\documentclass[12pt,review]{elsarticle}

\usepackage{url}
\usepackage{natbib}
\usepackage{amsmath,amssymb}
\usepackage{verbatim}
\usepackage{listings}
\usepackage{color,xcolor}
\def\urltilda{\kern -.15em\lower .7ex\hbox{\~{}}\kern .04em}

\journal{Spatial Statistics}

\begin{document}
\lstset{
language=R,
basicstyle=\scriptsize\ttfamily,
commentstyle=\ttfamily\color{blue},
stepnumber=1,
numbersep=5pt,
backgroundcolor=\color{lightgray},
showspaces=false,
showstringspaces=false,
showtabs=false,
frame=single,
tabsize=2,
captionpos=b,
breaklines=true,
breakatwhitespace=false,
escapeinside={},
keywordstyle={},
morekeywords={}
}

\begin{frontmatter}

\title{Detecting correlation between allele frequencies and environmental 
variables as a signature of selection.\\ A fast computational approach for genome-wide studies. }
\author[gg]{Gilles Guillot\corref{cor1}}
\author[rvmg]{Renaud Vitalis}
\author[alr]{Arnaud le {R}ouzic}
\author[rvmg]{Mathieu Gautier}

\cortext[cor1]{Corresponding author}

\address[gg]{Department of Applied Mathematics and Computer Science, \\
R. Pedersens Plads, 
  Technical University of Denmark, 2800 Kongens Lyngby, Denmark.}
\address[rvmg]{Centre de {B}iologie et de {G}estion des {P}opulations,
 34988 Montferrier-sur-Lez cedex, France}
\address[alr]{Laboratoire {\'E}volution, {G}\'enome et {S}p\'eciation, 91198, Gif-sur-Yvette, France}

\begin{abstract}
Genomic regions (or loci) displaying outstanding correlation with some environmental 
variables are likely to be under selection and this is the rationale of recent methods 
of identifying selected loci 
and retrieving functional information about them. 
To be efficient, such methods need to be able to disentangle the potential effect of environmental variables 
from the confounding effect of population history. 
For the routine analysis of genome-wide datasets, one also needs fast inference and model selection algorithms. 
We propose a method based on an explicit spatial model
which is an instance of spatial generalized linear
mixed model (SGLMM).
For inference, we make use of 
the INLA-SPDE theoretical and computational framework developed by \citet{Rue09} and \citet{Lindgren11}. 
The method we propose allows one to quantify the correlation between genotypes
and  environmental variables. 
It works for the most common types of genetic
markers, obtained either at the individual or at the population level. 
Analyzing simulated data produced under a
geostatistical model then under an explicit model of selection, we
show that the method is efficient. We also re-analyze a dataset
relative to nineteen pine weevils  ({\em  {H}ylobius abietis})
populations across Europe. The method proposed appears also as a 
statistically sound alternative to the Mantel tests for testing the
association between genetic and environmental variables. \\
\end{abstract}
\end{frontmatter}
\newpage

\section{Background}
\subsection{Detecting signature of natural selection}

Natural (or Darwinian) selection is the gradual process by which
biological traits (phenotypes) become either more or less common in a population as
a consequence of reproduction success of the individuals that bear them. 
 Over time, this process can result in populations that specialize for
 particular ecological niches and may eventually lead to the
 emergence of new species. 
The study of selection is an important aspect of evolutionary biology
as it provides insight about speciation but also about the genetic
response of possibly lesser magnitude to environmental variation. 
An important goal of such analyses consists in identifying
genes or genomic regions that have been the target of selection
\citep{Nielsen05,Hansen12}. 
Identifying such genes may provide important information about their function which may eventually
help improving crops \citep{Yamasaki13} and livestock \citep{Flori09}.
Recent genotyping techniques make it possible to obtain DNA sequences
at a high  number of genomic locations  in a growing number of both
model and non-model species \citep{Davey11}. This opens the door to
methods of identifying regions under selection, even for organisms whose genome is poorly
documented (non-model organisms), but the large size  of such datasets
($10^4$-$10^6$ variables) makes the task a formidable statistical challenge.

\subsection{Recent methods of detecting selection}
So far, identifying genomic regions targeted by selection
has relied extensively
on the analysis of genetic data alone, based on the idea that, if
local selection occurs at a given 
chromosome region (or locus),
differentiation (genetic difference between population) will increase at this
locus compared with what is theoretically expected at neutral loci
\citep{Nielsen05}.
To further identify the environmental characteristics
associated with the observed genetic variation, 
a  recent family of methods 
attempts to identify loci displaying outstanding correlation 
with some environmental variables. 
This more direct approach has the potential advantage to provide
functional information about those conspicuous loci. 
The data required for the latter type of analyses typically consist of the genotypes of
a set of individuals at various genomic loci and measurements of
various environmental variables at the same sampling sites.  The method amounts to quantifying
the statistical dependence between allele counts and environmental
variables.

The most natural method to model  dependence of count
data on a quantitative or qualitative variable is the logistic
regression, as implemented in this context by \citet{Joost07}. 
However, plain logistic regression  assumes that allele
counts among different populations or individuals are independent
conditionally on the environmental variable. 
Doing so, logistic regression fails to capture the residual genetic
dependence of neighboring individuals or populations due to their
common  ancestry and recent common evolutionary history. 
Another method to test the dependence between genetic 
and environmental variables is the Mantel test and its variant the
partial Mantel test. These tests attempt to assess the significance 
of a correlation coefficient by re-sampling with permutations. 
They have long been popular methods in ecology and evolution. 
However, a recent study \citep{Guillot13} show that they  are not appropriate
if the data are spatially correlated. 
The method proposed by \citet{Coop10} attempts to model  genetic structure 
by including a random term in the logistic regression in a fashion 
similar to a Generalized Linear Mixed Model. They propose to do
inference with MCMC. 
A recent study by  \citet{deMita13} shows that under biologically
realistic conditions, accounting for structure in the data as in
\citep{Coop10} improves the accuracy of inferences. 
The goal of the present paper is to extend the latter approach by
rooting it in a spatially explicit model and implementing inference with an MCMC-free
inference approach.

The method proposed is described in the next section. Next we
illustrate the method accuracy by analyzing simulated data
produced first under a purely geostatistical model then under a biological
model that simulates selection explicitly. We conclude by discussing
our results and outlining possible extensions.

\section{Method proposed}
\subsection{Data considered}
We consider a set of individuals observed at various 
geographical locations. 
Each individual is genotyped at $L$ genetic loci. Besides, 
we consider that these loci are bi-allelic, 
i.e the sequence observed at
a particular locus can be only of two types (denoted arbitrarily A/a in the sequel). 
We consider haploid or diploid organisms, i.e. organisms that carry either one or two copies of each
chromosome. A genotype  is therefore a vector with $L$
entries in $\{0,1\}$ or $\{0,1,2\}$ respectively. 
As it is frequent to sample more than one individual at each location, 
we denote by  $n_{il}$ the haploid sample size of population $i$ for
locus $l$, that is 
the number of individuals at site $i$ genotyped at locus $l$ times the number of chromosome copies 
carried by the organism under study.

\subsection{The pine weevil dataset}
\begin{figure}[h]
\begin{center}
\vspace{-1cm}\includegraphics[width=13cm]{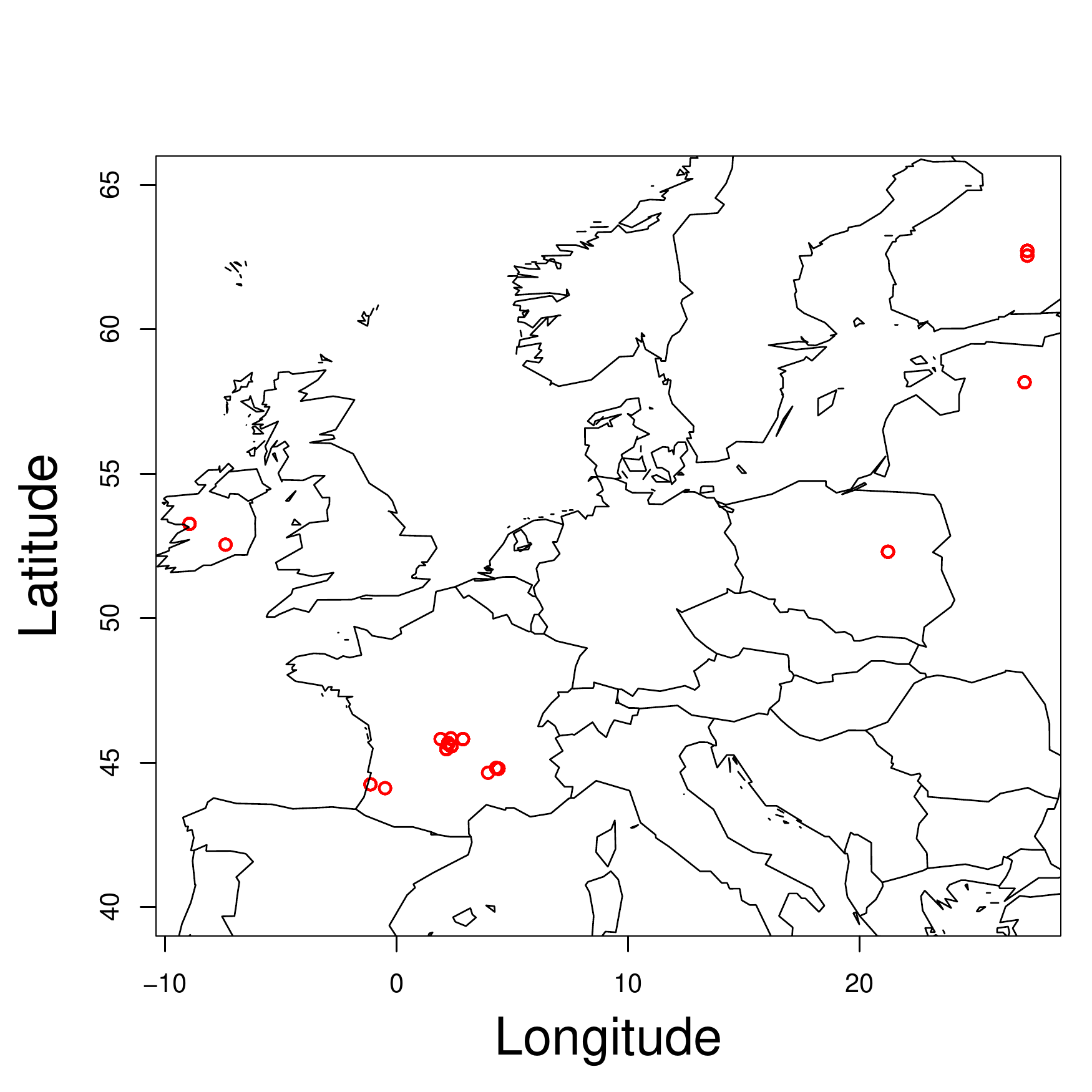}
\end{center}
\vspace{-1cm}\caption{Geographical locations of the nineteen pine
  weevil populations sampled in Europe.}\label{fig:weevil_coord}
\end{figure}

To illustrate the method proposed here, 
we will re-analyse a dataset relative to pine weevils 
initially produced by \citet{Conord06}. 
Anticipating on the results section and for the sake of fleshing out the
presentation of the method in the next section, we briefly outline the main
features of this dataset.
It consists of 367 pine weevil individuals  ({\em  {H}ylobius abietis}) 
sampled in 19 geographical locations across Europe (figure
\ref{fig:weevil_coord}). 
Each individual has been genotyped at 83 genetic markers 
(see below for details). 

This dataset has been analysed by \citet{Joost07} who looked for
signatures of selection by comparing spatial genetic variation to
ten  environmental variables. 
We focus here on a subset consisting of four environmental
variables: 
average diurnal temperature range, number of days with ground frost,
average monthly precipitation and average wind speed. 

\subsection{Model}

\subsubsection{Likelihood}
We denote by $(s_i)_{i=1,...,I}$ a collection of geographical
coordinates, $(y_i)_{i=1,...,I}$ some measurements of an environmental
variable obtained at  these sites and 
$(z_{il})_{\stackrel{i=1,...,I}{l=1,...,L}}$ the number of alleles of type A at locus $l$
observed in a population sampled at site $i$. 
We also denote by $f_{il}$  the local frequency of allele A at geographical
location $s_i$ for locus $l$. 
We make the assumption that there is no within-population statistical
structure and that for organisms harboring more than one copy of each
chromosome, the various alleles carried at a locus by an individual
are independent 
and that the allele counts are sampled from a binomial distribution:
\begin{equation}\label{eq:binom}
z_{il} \sim  \mbox{Binom}(n_{il},f_{il})
\end{equation} 
where $n_{il}$ denotes the number of alleles sampled (or haploid
 sample size) at site $i$. 

The above assumes that the data at hand provide exact information
about the  alleles carried by each individual. This is not the case
for certain genetic markers such as amplified fragment length
polymorphism markers (AFLP). 
With this type of markers, one can only know whether an individual
carries allele $A$ or not but the number of copies carried by each
individual is not known. 
For diploid organisms, this leads to a genotype ambiguity: 
the record of allele $A$  may correspond to genotypes $(a,A)$ or $(A,A)$. 
We therefore consider an alternative likelihood for the case above
where $z_{il}$ denotes the number of individuals at sampling site $i$ 
for which allele $A$ has been observed. 
Still denoting by $f_{il}$  the  frequency of a reference allele $A$ at
locus $l$ at geographical site $i$ but we have now 
\begin{equation}\label{eq:clump_binom}
z_{il} \sim  \mbox{Binom}(n_{il},f'_{il}) 
\mbox{\;\;\;\;\; with \;\;\;\;\;} f'_{il}=2f_{il}(1-f_{il})+f_{il}^2
\end{equation} 

\subsubsection{Latent Gaussian structure}

We model the dependency between an environmental variable $y$ and 
the allele frequency at locus $l$ by assuming that 
\begin{equation}\label{eq:logit}
f_{il}  =  \frac{1}{1+\exp - (x_{il} + a_l y_{i} + b_l)}
\end{equation}
where $x_{il}$ is an unobserved spatially random effect  that accounts for spatial auto-correlation due 
to population history and $(a_l,b_l)$ are parameters that quantify the locus-specific
effect of the environment variable $y_i$. 
The environment variable is observed and is treated as a spatially
variable explanatory variable (fixed effect).

The variables ${\bf x}_l = (x_{1l},...,x_{Il})$ are un-observed random effects and  are assumed to be
independent replicates from the same Gaussian random field. 
Doing so, we assume the absence of linkage disequilibrium (i.e absence
of statistical dependence across loci). 
By assuming a common distribution for all vectors ${\bf x}_l$, 
we inject the key information in the model that there 
is a characteristic spatial scale that is common to all loci and reflects 
the species- and area-specific population structure of the data  under study.

As commonly done in spatial statistics \citep{Chiles99}, we make the assumption that $x$ is 
0-mean isotropic and stationary.
Further, we assume that the stationary covariance $C(s,s')=C(h)$ belongs to the Mat{\'e}rn family i.e. 
\begin{equation}
  C(h) = \frac{\sigma^2}{2^{\nu-1} \Gamma(\nu)}(\kappa h)^{\nu}K_{\nu}(\kappa h)
\end{equation}
where $K_{\nu}$ is the modified Bessel function of the second kind and order 
$\nu>0$, 
$\kappa>0$ is a scaling parameter 
and $\sigma^2$ is the marginal variance.

\subsection{Parameter inference} 
A key feature of the model above is that it can be handled within
the theoretical and computational framework  developed by \citet{Rue09} and 
\citet{Lindgren11}. 
The former develops a framework for Bayesian inference in a broad class of models enjoying 
a latent Gaussian structure. 
The latter, bridges a gap between Markov random fields and Gaussian
random fields theory making it possible to combine the flexibility of Gaussian
random fields for modelling  and the computational efficiency of 
Markov random fields for inference. 
The approach of \citet{Lindgren11} is based on the observation that a Gaussian random field
$x(s)$ 
with a Mat{\'e}rn covariance function is the solution of the stochastic
partial differential equation 
\begin{equation}\label{eq:spde}
(\kappa^2 -\Delta)^{\nu /2}(\tau x(s)) = {\cal W}(s)
\end{equation}
where $\Delta$ is the Laplacian, $\kappa$ is the scale parameter,
$\nu$ controls the smoothness and $\tau$ controls the variance. 
In approximating $x(s)$ by 
\begin{equation}
x(s) = \sum_{k} \psi_k(s) w_k
\end{equation}
where the $\psi_k(.)$  are basis functions with compact
support, 
 one can choose the weights $w_k$ so that the distribution of the
 function $x(s)$ approximates the distribution of  the solution of
 Eq.~\ref{eq:spde}. \\
The method of \citet{Rue09} is based on Laplace approximations of
the various conditional densities involved in the inference of
the hyper-parameters and latent variables. It makes use of the Markov
structure of the latent variables in the computation. 
In contrast with MCMC, the INLA method does not compute estimates of the joint
posterior distributions of hyper-parameters and latent
variables  but it only 
estimates the marginal posterior densities.

Casting the present problem in the framework of INLA-SPDE   
opens the door to accurate and  fast computations 
using the R  package {\tt inla}. 
For parameter inference, data relative to all loci are combined 
 into a matrix $Z=({\bf z}_l)_{l=1,...,L}$ to compute 
the marginal posterior distribution $\pi(\kappa|Z)$ and
$\pi(\sigma^2|Z)$. 
For computational reasons \citep{Lindgren11},  the smoothness parameter 
$\nu$ is taken equal to one. A log-Gamma prior is assumed for $\kappa$
and a Normal prior is assumed for the fixed effect ($a_l$ and $b_l$). 
From these marginal posterior distributions, we derive estimates of
$\kappa$ and $\sigma$ as posterior means. 
In presence of a large number of loci, implementing the above 
on the full dataset may become unpractical due to memory load
issues. In this case we recommend to infer the parameters of the
spatial covariance on a random subset of loci. 

\subsection{Model selection}
For each locus, we are concerned with selecting among two competing
models: 
a model in which the environment has an effect, i.e. where 
$f_{il}  =  1/[1+\exp - (x_{il} + a_l y_{i} + b_l)]$
and a reduced
model in which the environmental variable has no effect, namely $a_l=0$
in the previous equation.
For the vector ${\bf z}_l=(z_{1l},...,z_{Il})$ of data at locus $l$, 
we denote  $ \displaystyle \pi({\bf z}_l|m) = \int
\pi({\bf z}_l|\theta,m)\pi(\theta|m)d\theta $ the evidence or integrated
likelihood of data under model $m$.
Assessing the strength of association with environmental
variables of selection can be done
by  computing the Bayes factor  
\begin{equation}
BF_l = \pi({\bf z}_l|m_1) / \pi({\bf z}_l|m_0)
\end{equation}

We compute Bayes factors
$BF_l$ and estimate  $a_l$
and $b_l$ locus-by-locus. In this second step of computations, 
the variance and scale parameters are fixed to the values inferred
from the global dataset $Z$ as explained in the previous section.  
The Bayes factors  can be used to flag loci displaying
outstanding dependence with the environmental variables and rank loci
by decreasing evidence of genetic selection.

\section{Analysis of simulated and real data}

\subsection{Simulations from a geostatistical model}

\begin{figure}[h]
\vspace{-.0cm}\begin{tabular}{cc}
\vspace{-.2cm}\includegraphics[width=7.8cm]{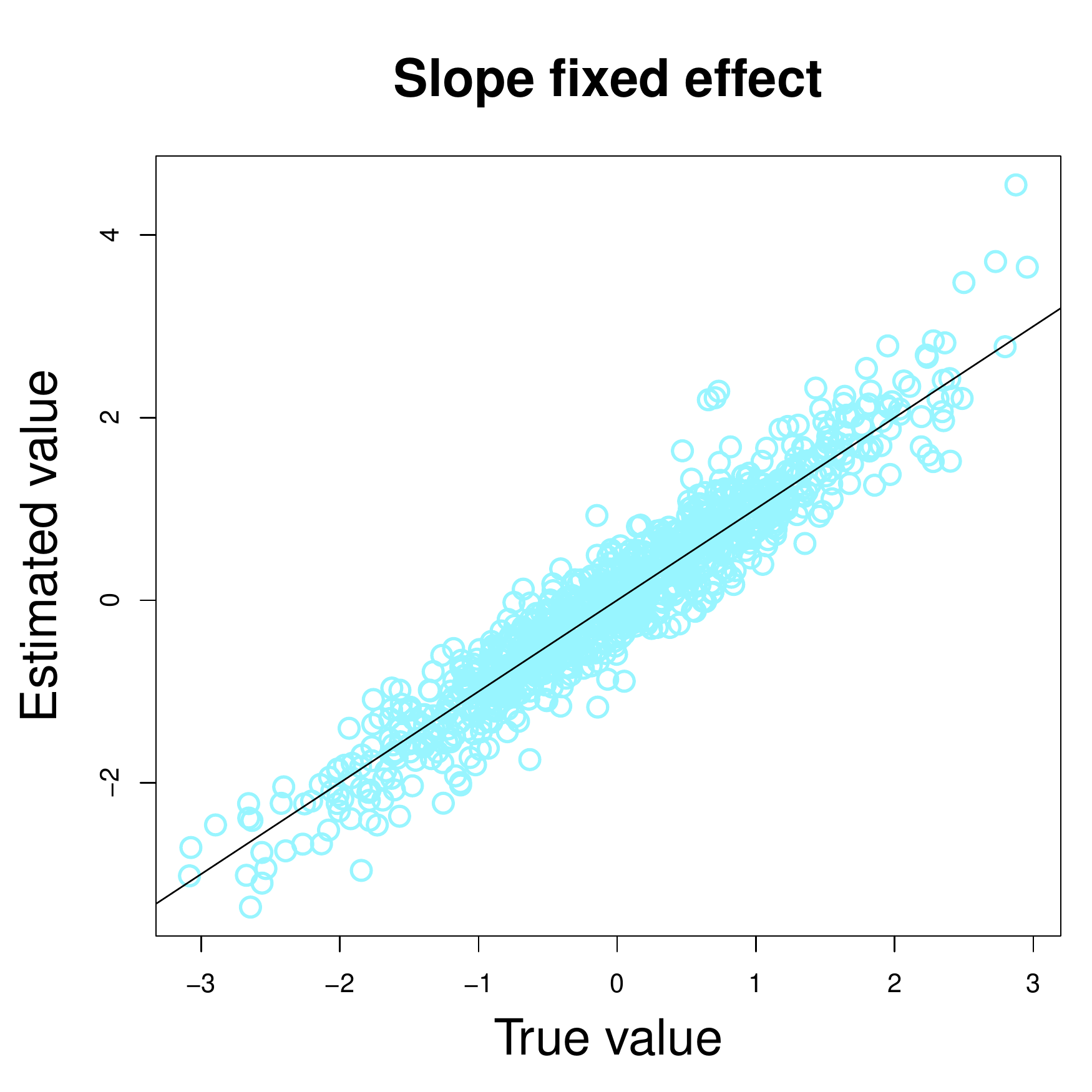}
& \vspace{-.2cm}\includegraphics[width=7.8cm]{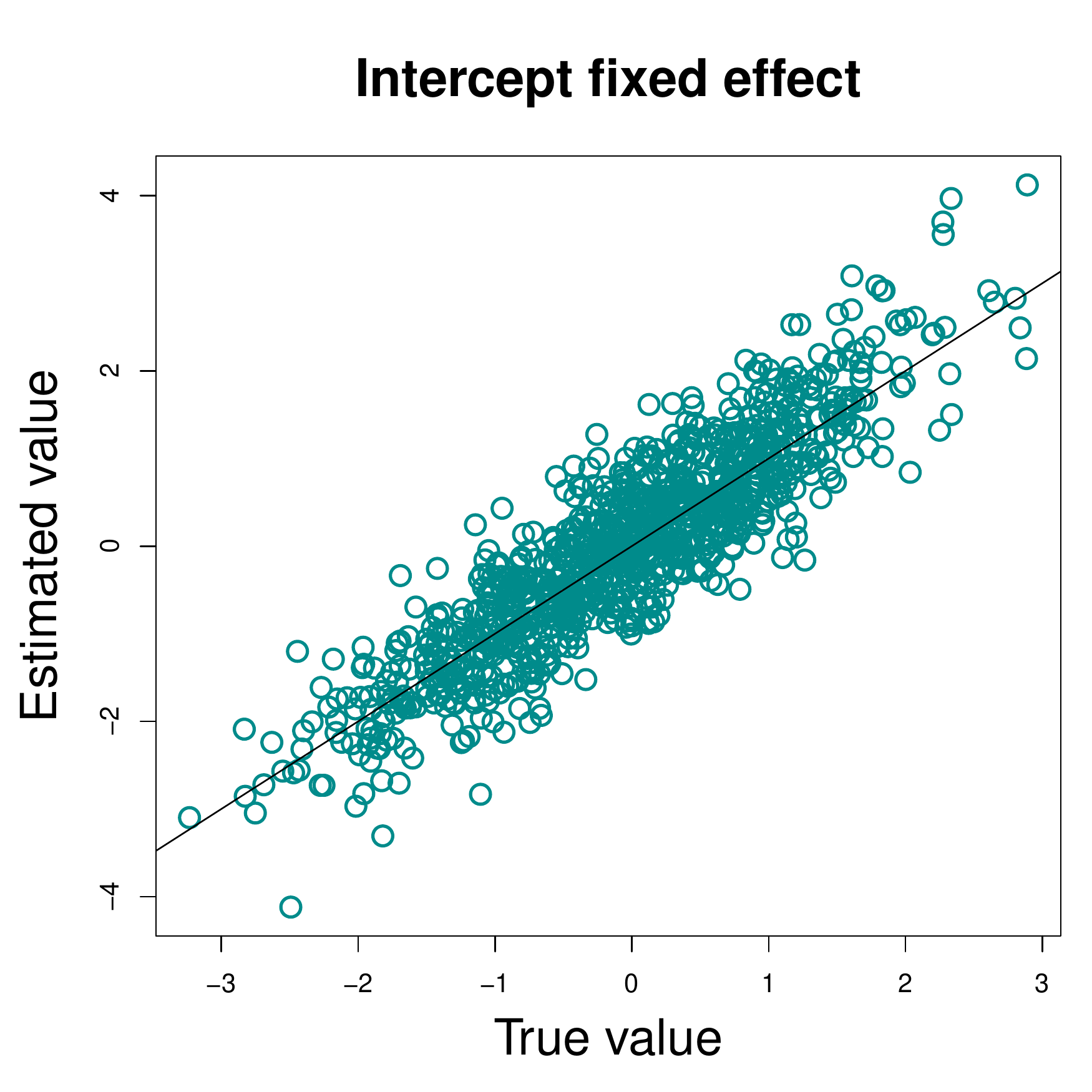} \\
\vspace{-.2cm}\includegraphics[width=7.8cm]{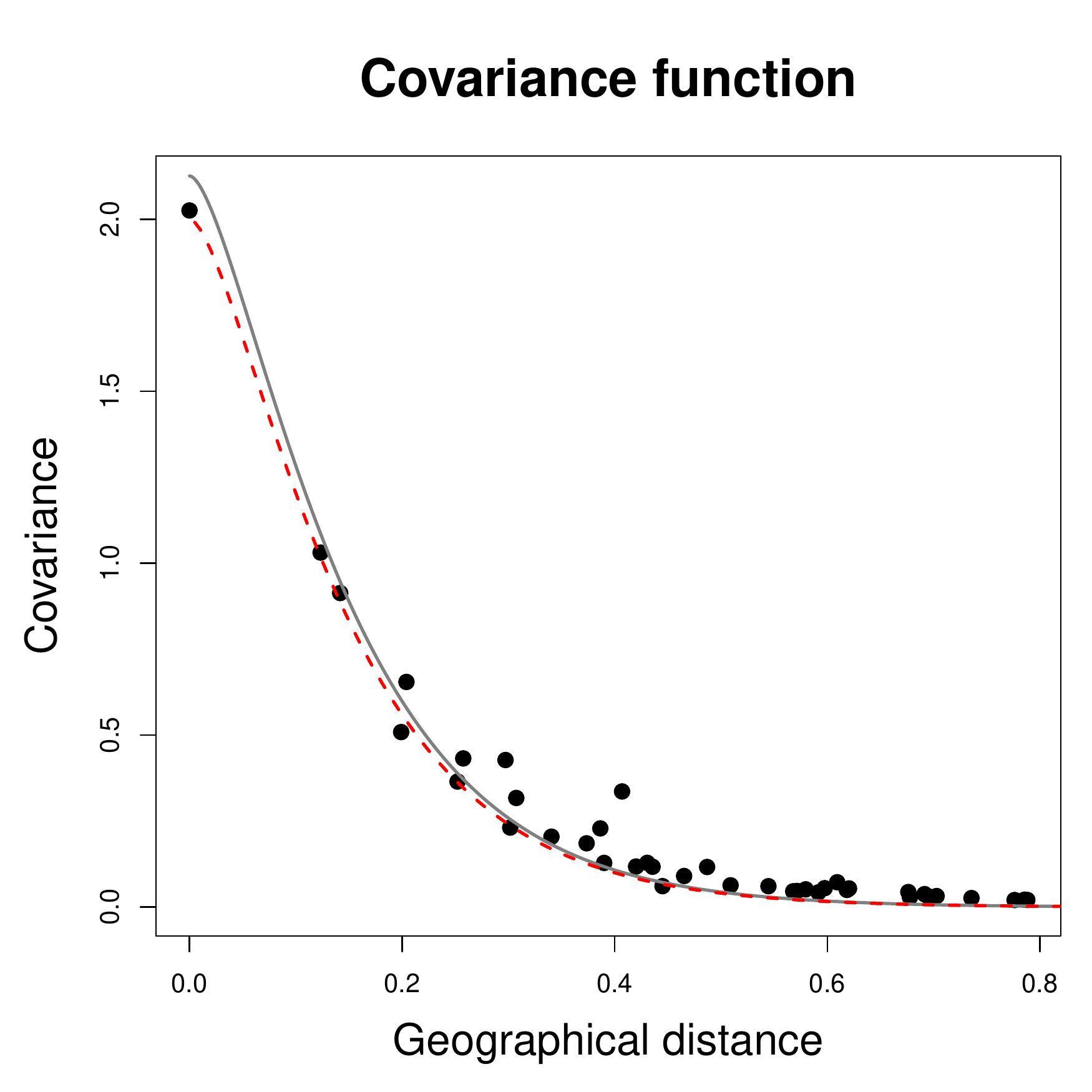}
& \vspace{-.2cm}\includegraphics[width=7.8cm]{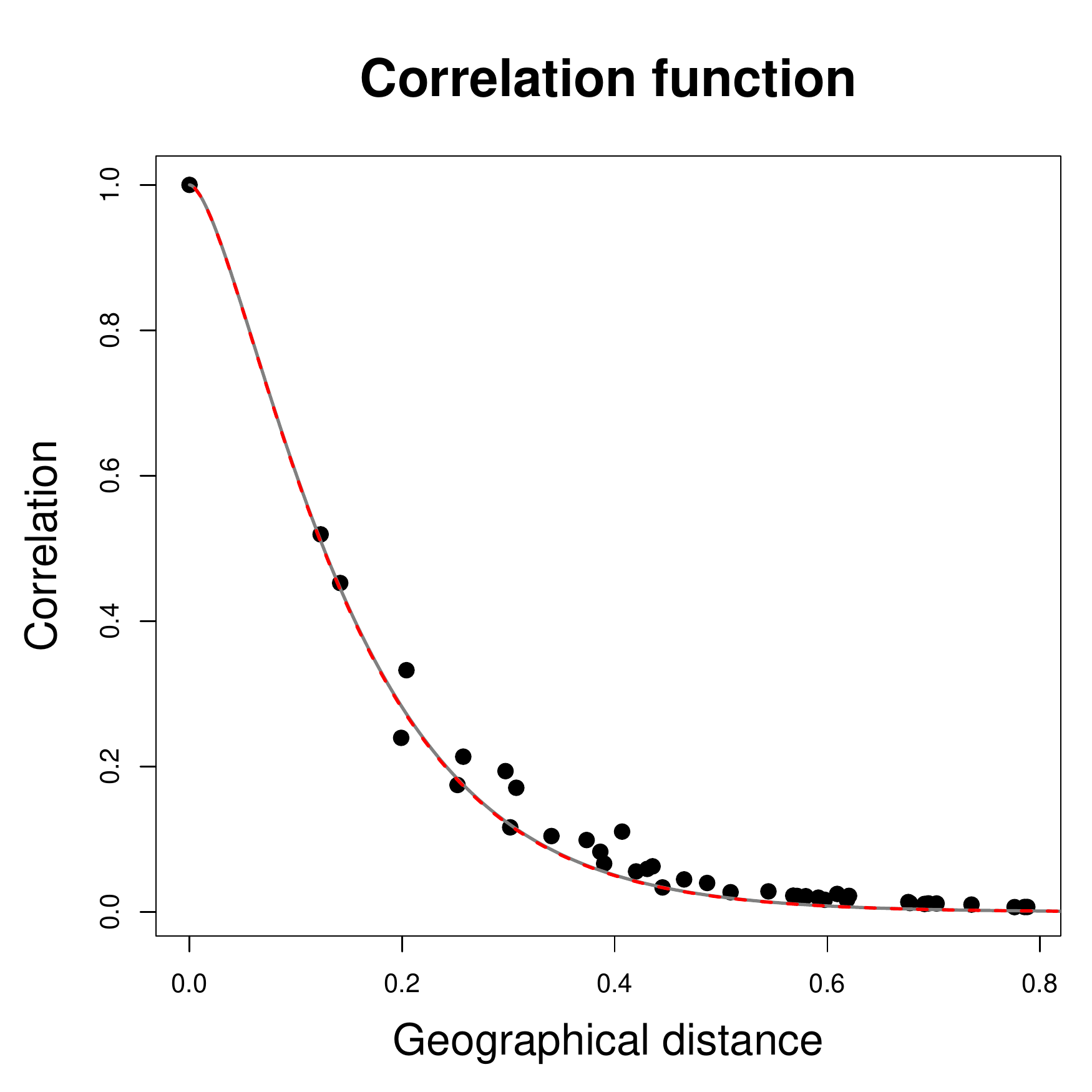}
\end{tabular}
\caption{Results of inference on data  from geostatistical
  simulations. 25 geographical sites, 20 individuals per sites, 1000
  AFLP markers. Top row: slope $a_l$ and intercept $b_l$ of fixed
  effect (Eq.~\ref{eq:logit}). 
Bottom row: the dashed red lines depicts the true
  Mat{\'e}rn   covariance and correlation functions for the hidden Gaussian fields, the
  continuous grey line depicts the estimated  Mat{\'e}rn functions and the black
  dots the numerical result for the GMRF approximation underlying the
  INLA-SPDE method.}\label{fig:geostat_R2_AFLP}
\end{figure}

We analyse here data simulated under the exact model described above. 
We consider first a dataset of 1000 bi-allelic dominant markers
(Eq. \ref{eq:clump_binom}) 
for 500 individuals observed at 25 geographical sites
uniformly sampled in the unit square (20 individuals per site),
which  are typical sample sizes encountered in molecular ecology
studies. 
For the fixed effects (Eq. \ref{eq:logit}), we draw 
$a_l$, $b_l$ and $(y_i)_{i=1,...,I}$ independently from a $N(0,1)$ distribution. 
The random effect ${\bf x}_l$ is a Gaussian random field with a
Mat{\'e}rn covariance function with parameters $\sigma^2=2$, $\nu=1$ and
$\kappa=0.1$. 
The results of inference reported
figure \ref{fig:geostat_R2_AFLP} show an excellent accuracy in the
inference of the underlying covariance function and also a good
accuracy in the estimation of the fixed effect (parameters $a_l$ and
$b_l$). 
In the inference with INLA,  we use everywhere the default prior
distributions. 
In other simulation experiments under the same geostatistical model
with other combinations of parameters,  we observed
sometimes that the estimation of the variance parameter could be
inaccurate. 
For example, with a range parameter $\kappa=0.1$ 
However, this does not seem to affect the accuracy in the
estimation of the other parameters. In particular, the slope $a_l$ in the
fixed effect which quantifies the effect of the environmental variable is consistently
accurately estimated. 

\subsection{Simulations from a landscape genetic model}
Individual-based simulations are produced here using the computer program
SimAdapt \citep{Rebaudo13}. 
The genome of each individual consists in 120 genetically-independent
bi-allelic co-dominant markers: 100 neutral loci  and 20 loci under
habitat-specific selection. Alleles at non-neutral loci are specific
to one of the two habitats 1 or 2. 
Homozygotes (A,A) in habitat 1 have a fitness of 1, while homozygotes
(a,a) have a fitness of 
$1-s$ (and vice versa in habitat 2). 
The fitness of heterozygotes is $1-s/2$ in both habitats.
Locus-specific fitnesses combine multiplicatively across loci to give
the fitness of individuals. 
Among the selected loci, ten are subject to selection in
one habitat, the ten others in the other habitat. 
Individuals are considered as hermaphrodites, and mate randomly in
their patch, the mating probability being proportional to their
fitness. Additional details on the model are provided in \citet{Rebaudo13}.

\begin{figure}[h]
\vspace{-1.5cm}\begin{tabular}{cc}
\vspace{-.5cm}\includegraphics[width=7.5cm]{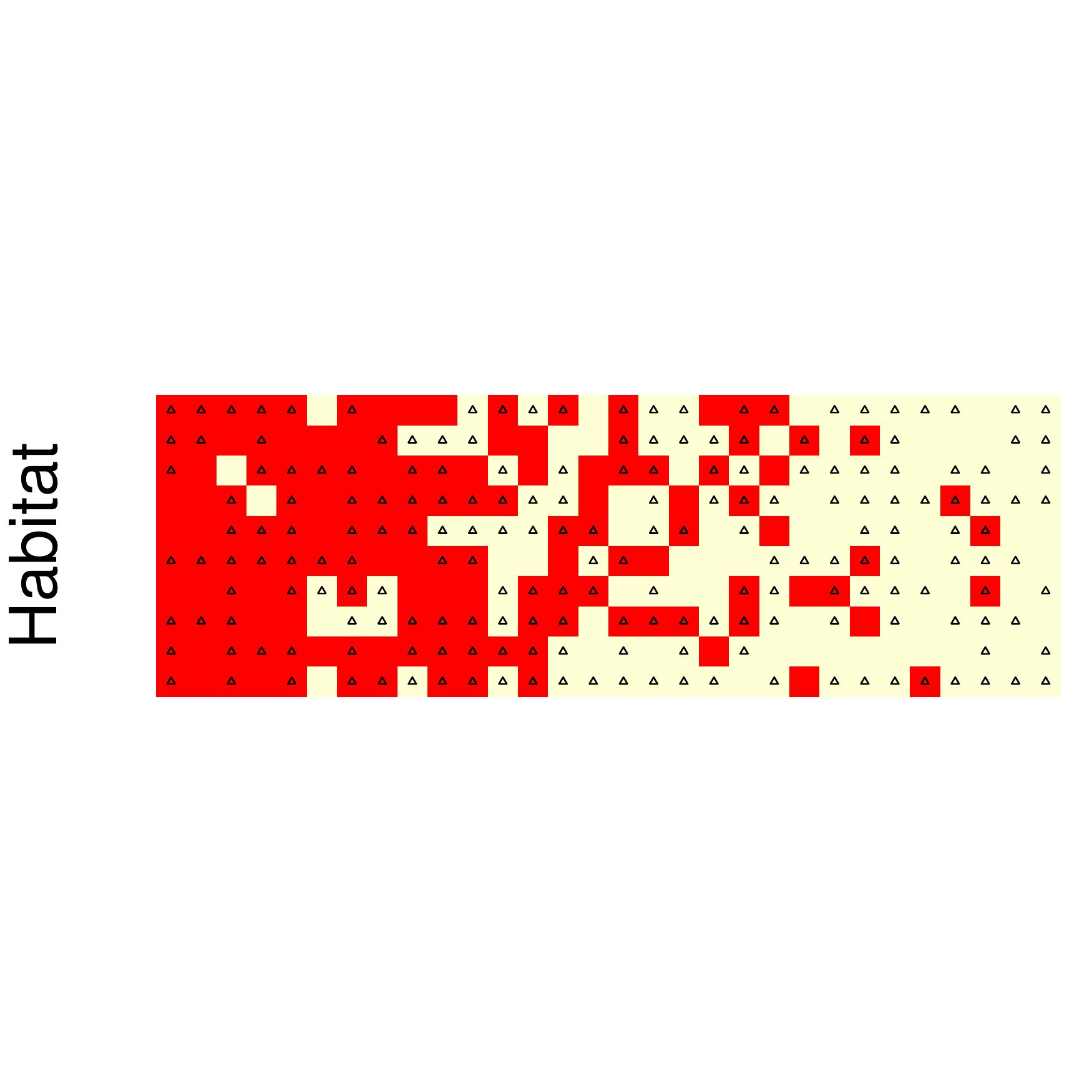}
& \vspace{-.5cm}\includegraphics[width=7.5cm]{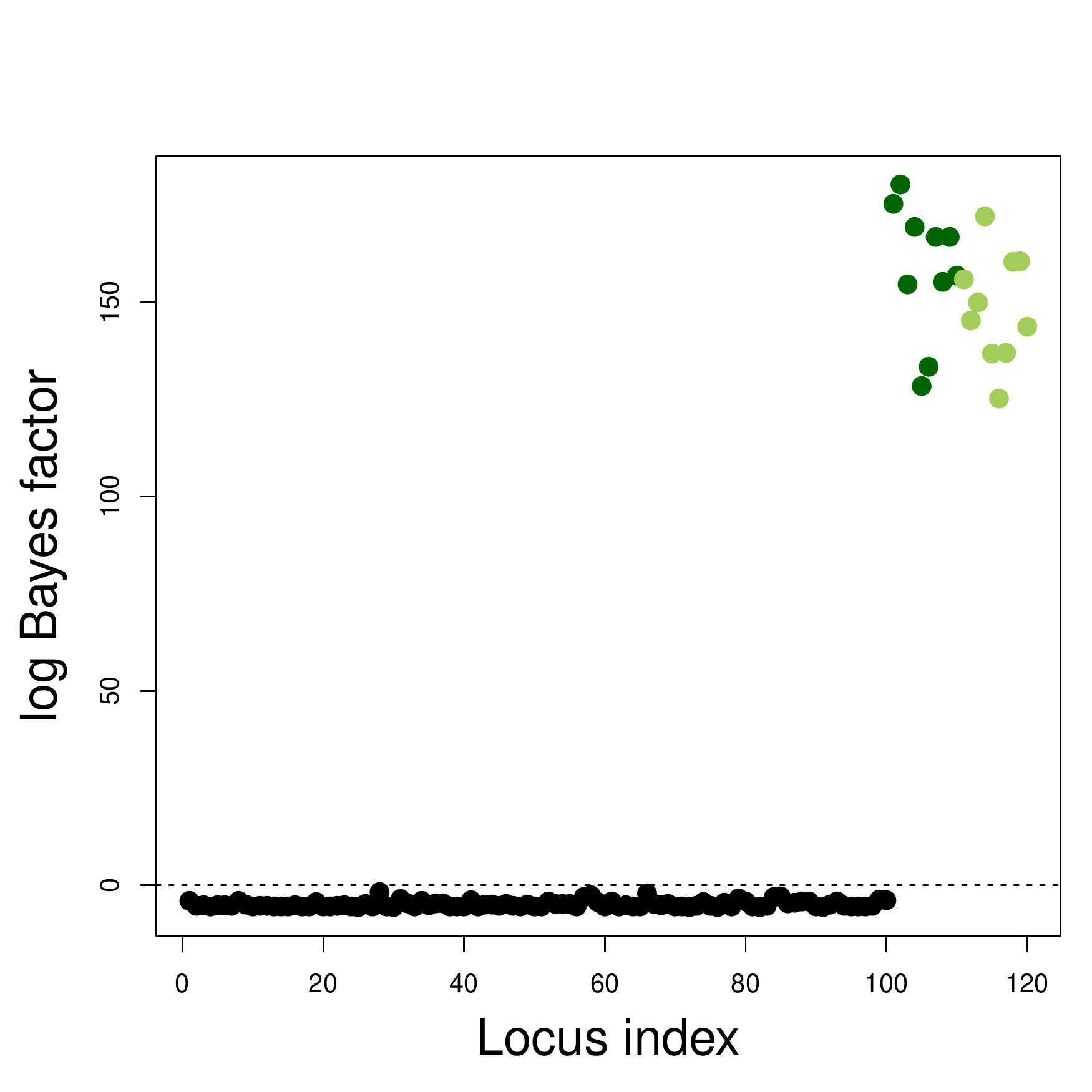} \\
\vspace{-.5cm}\includegraphics[width=7.5cm]{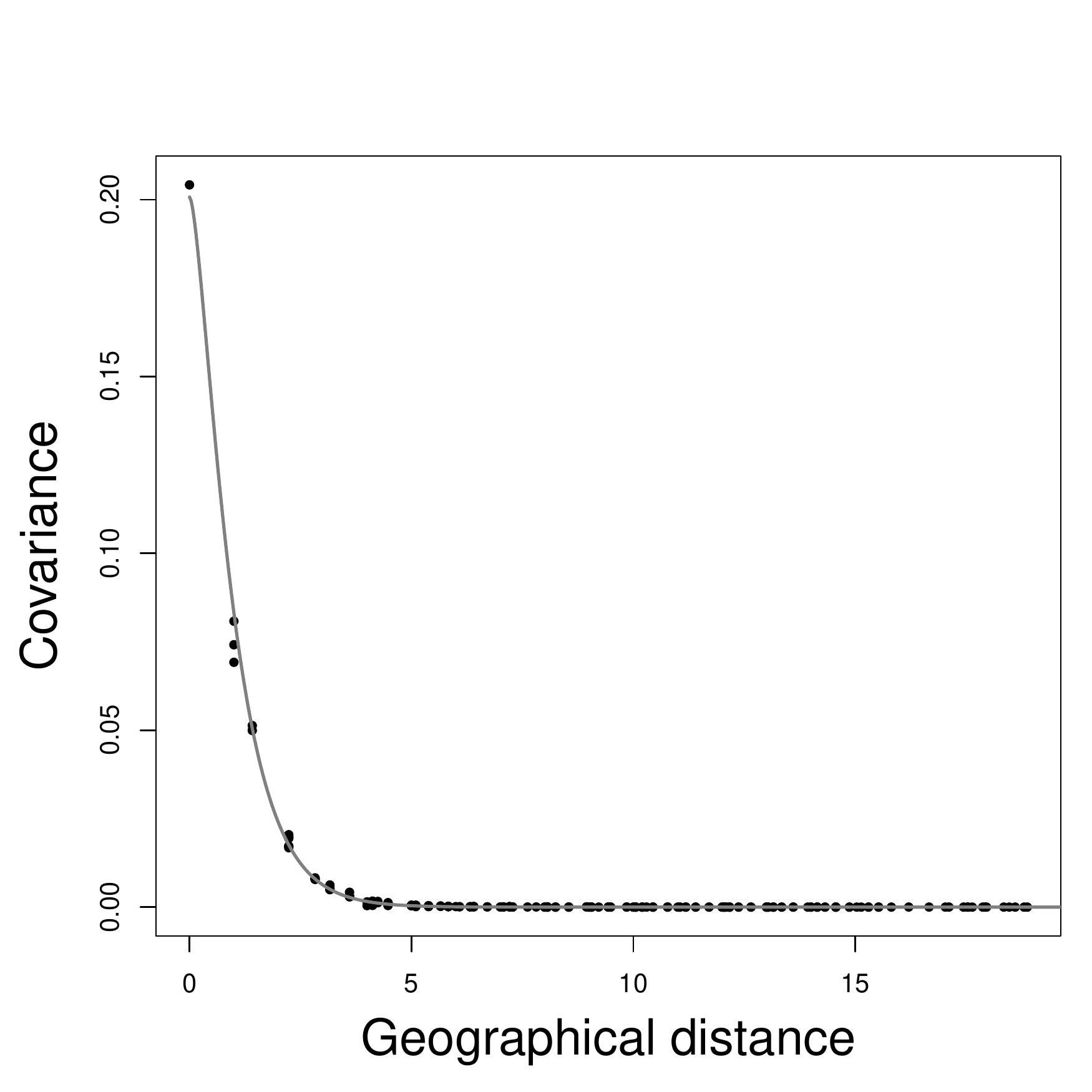}
& \vspace{-.5cm}\includegraphics[width=7.5cm]{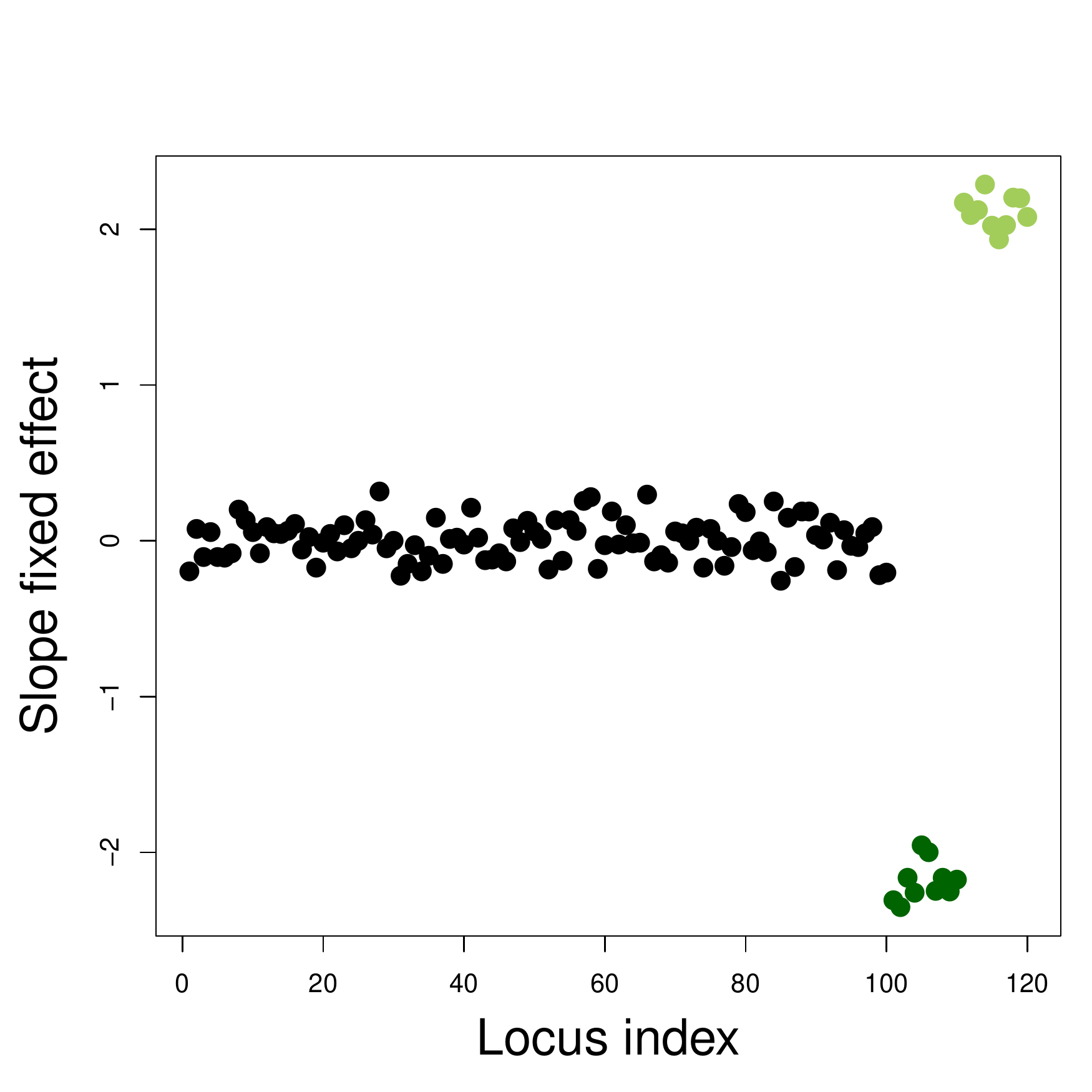}\\
 \includegraphics[width=7.5cm]{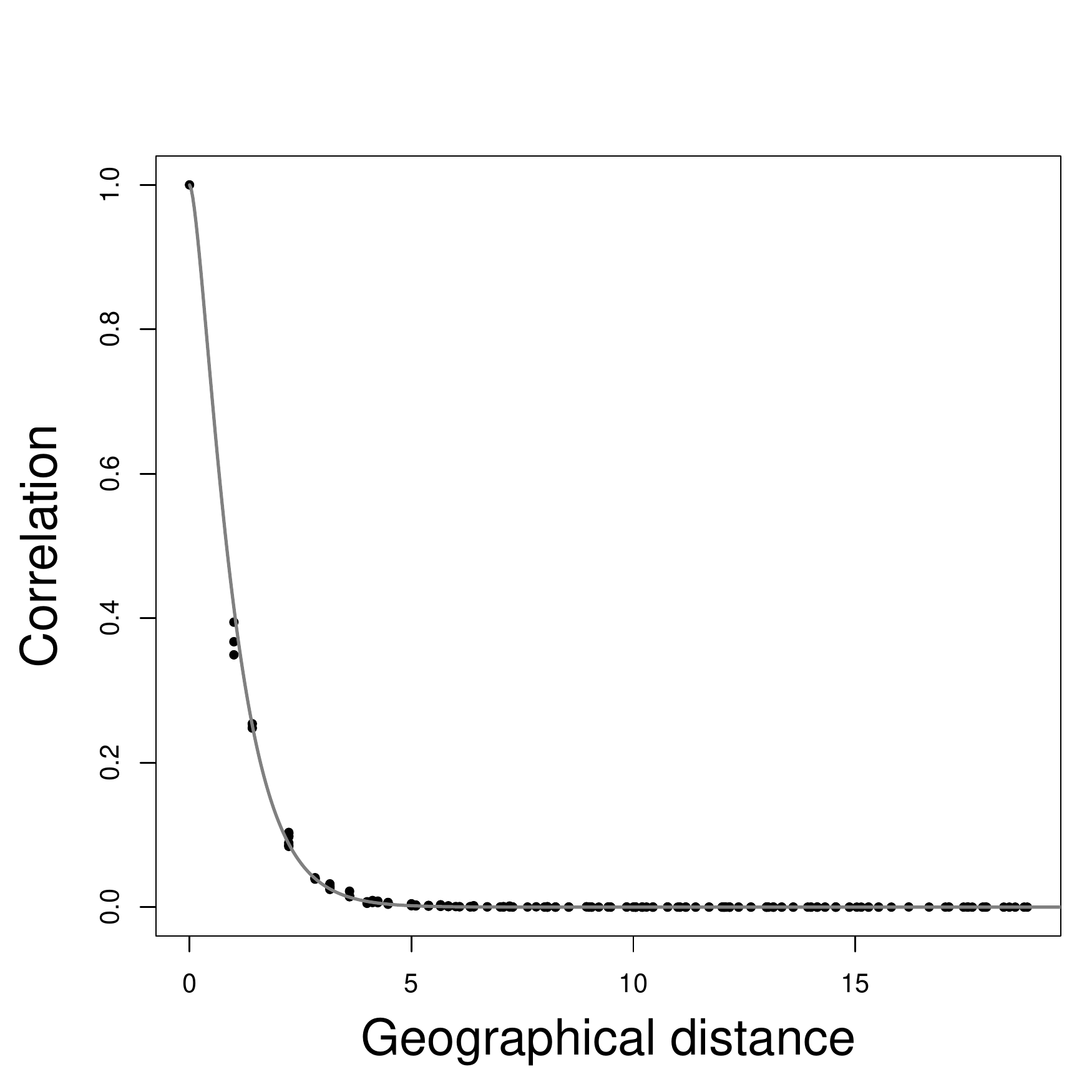} 
& \includegraphics[width=7.5cm]{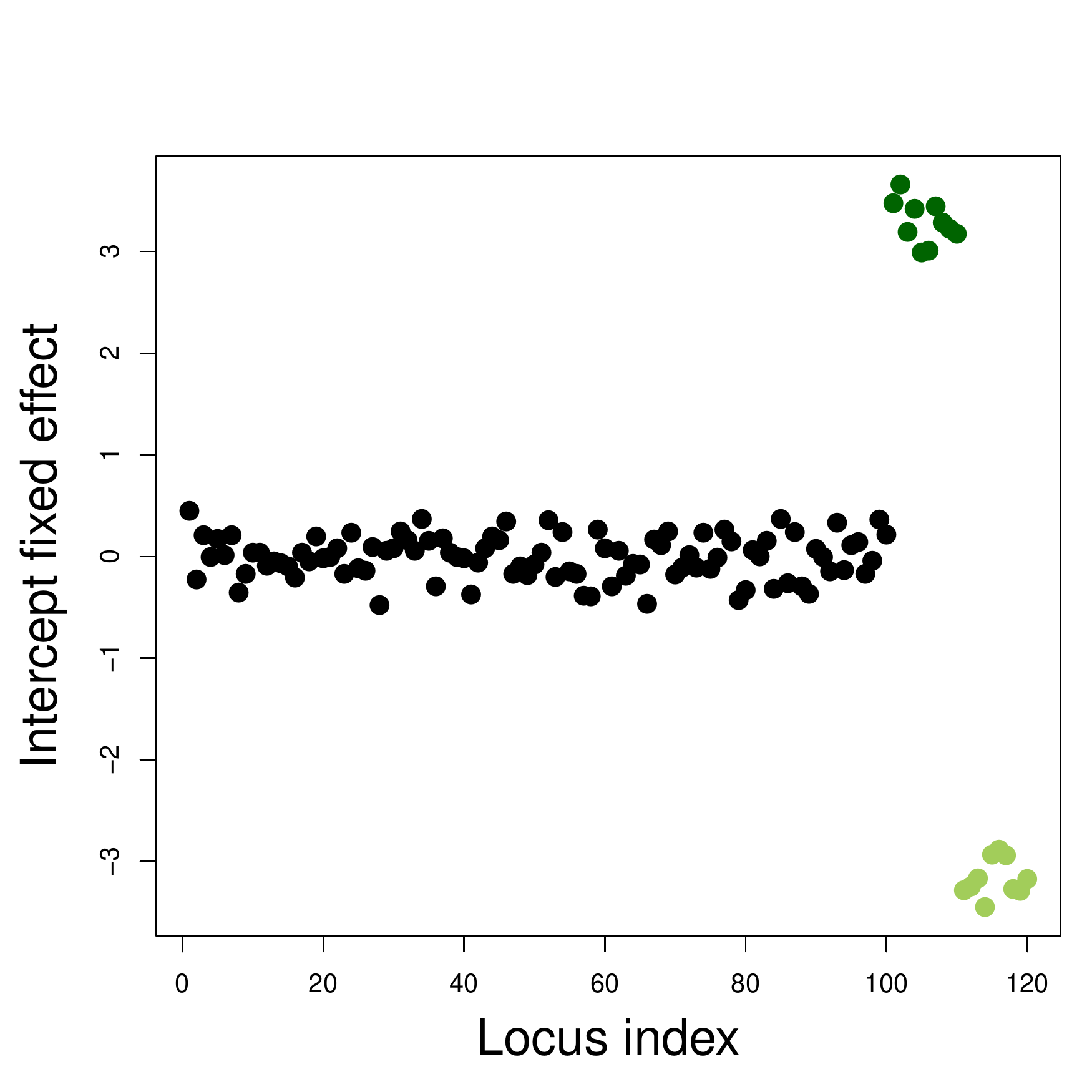}
\end{tabular}
\caption{
Results for data
  simulated from a landscape genetics model (dispersal
  probability=$0.1$ per individual and per generation). 
Top left: habitat (environmental variable) coded as two colors and
sampling sites (triangles); Middle left 
  and bottom left: the   continuous grey line depicts the estimated  Mat{\'e}rn functions and the black
  dots the numerical result for the GMRF approximation underlying the
  INLA-SPDE method. 
Right from top to bottom: Bayes factors and parameters $a_l$ and $b_l$ for the 120 loci. 
Dark and light green correspond to positively and
negatively selected loci respectively. 
The loci genuinely under selection are   indexed 101-120. }\label{fig:d01}
\end{figure}

\begin{figure}[h]
\vspace{-1.5cm}\begin{tabular}{cc}
\vspace{-.5cm}\includegraphics[width=7.5cm]{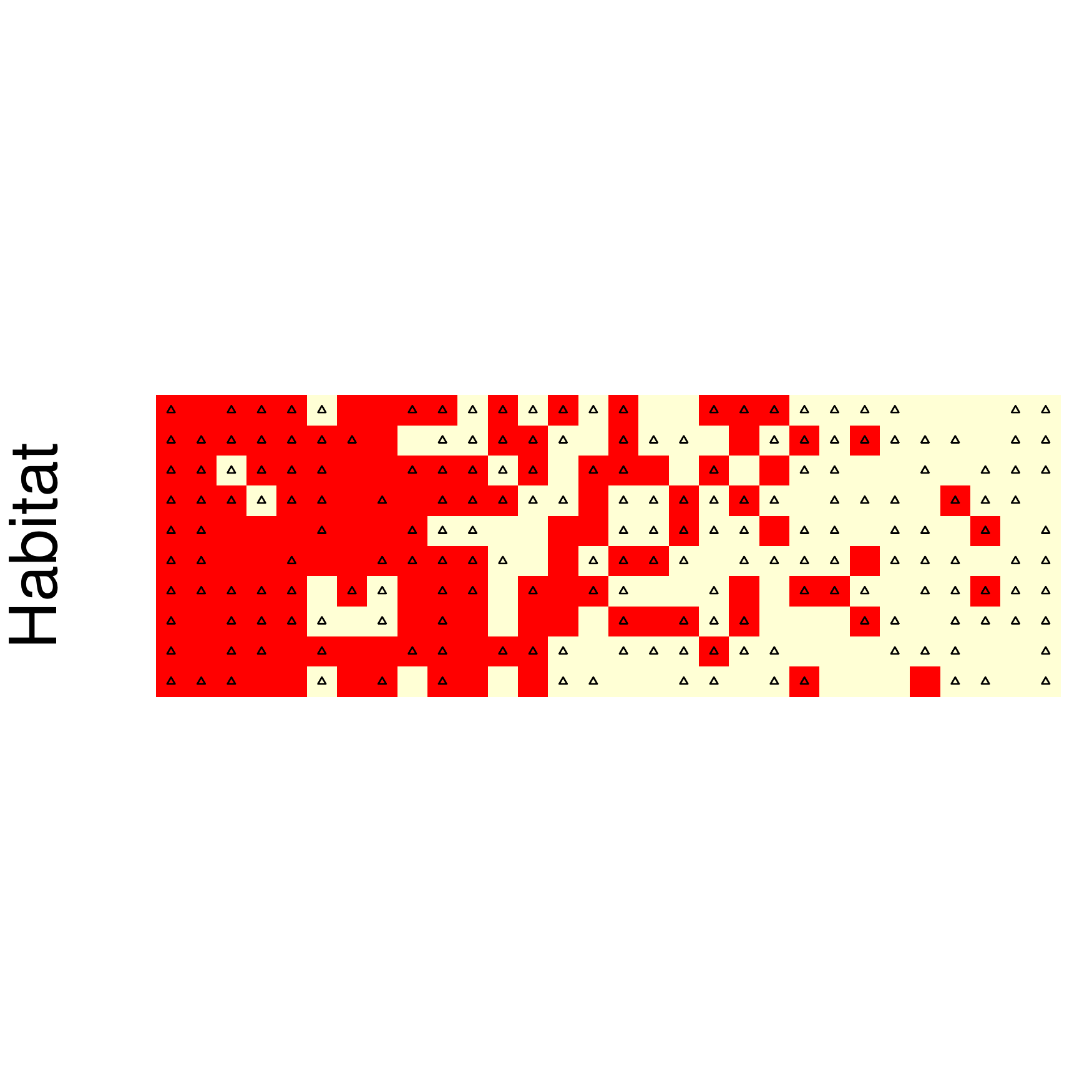}
& \vspace{-.5cm}\includegraphics[width=7.5cm]{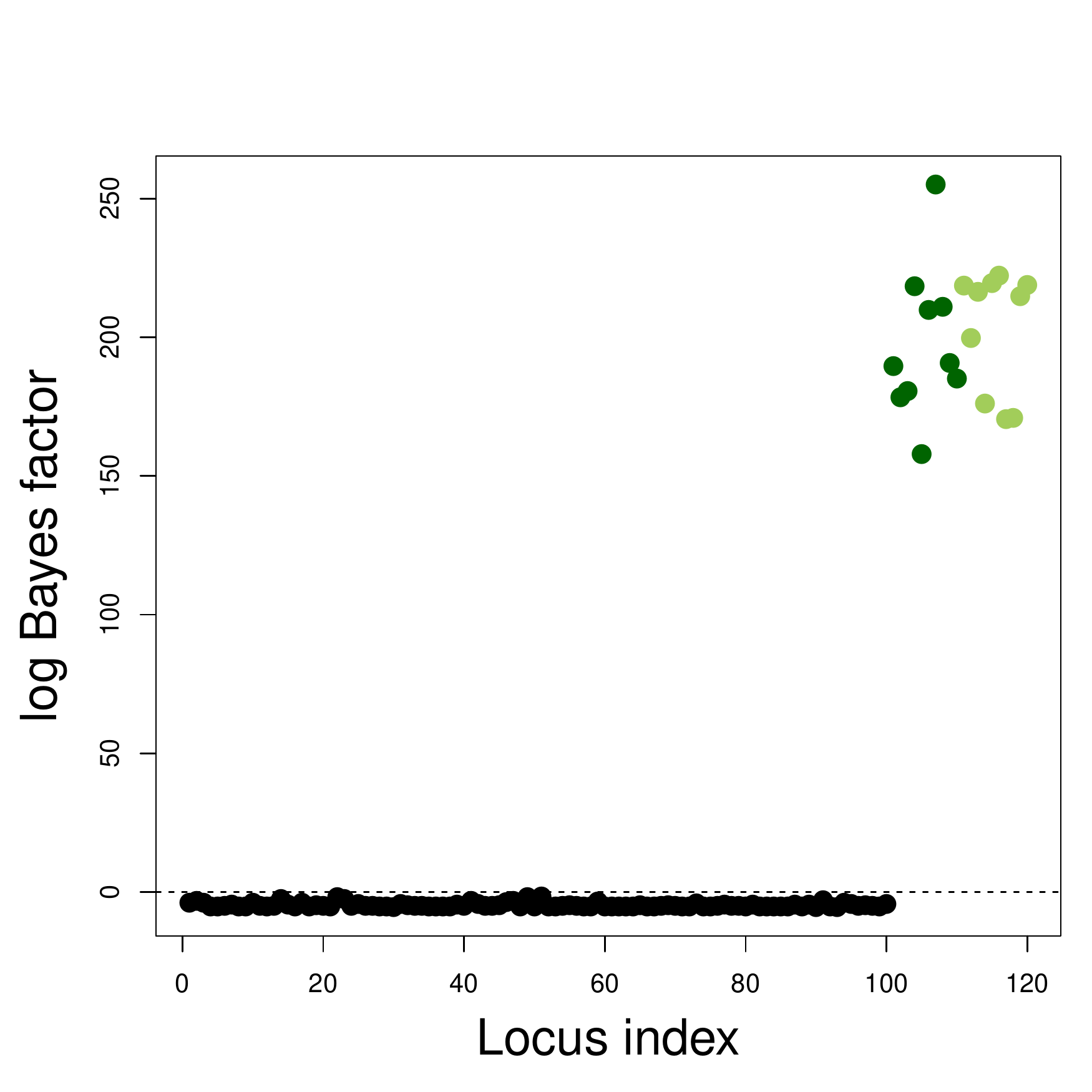} \\
\vspace{-.5cm}\includegraphics[width=7.5cm]{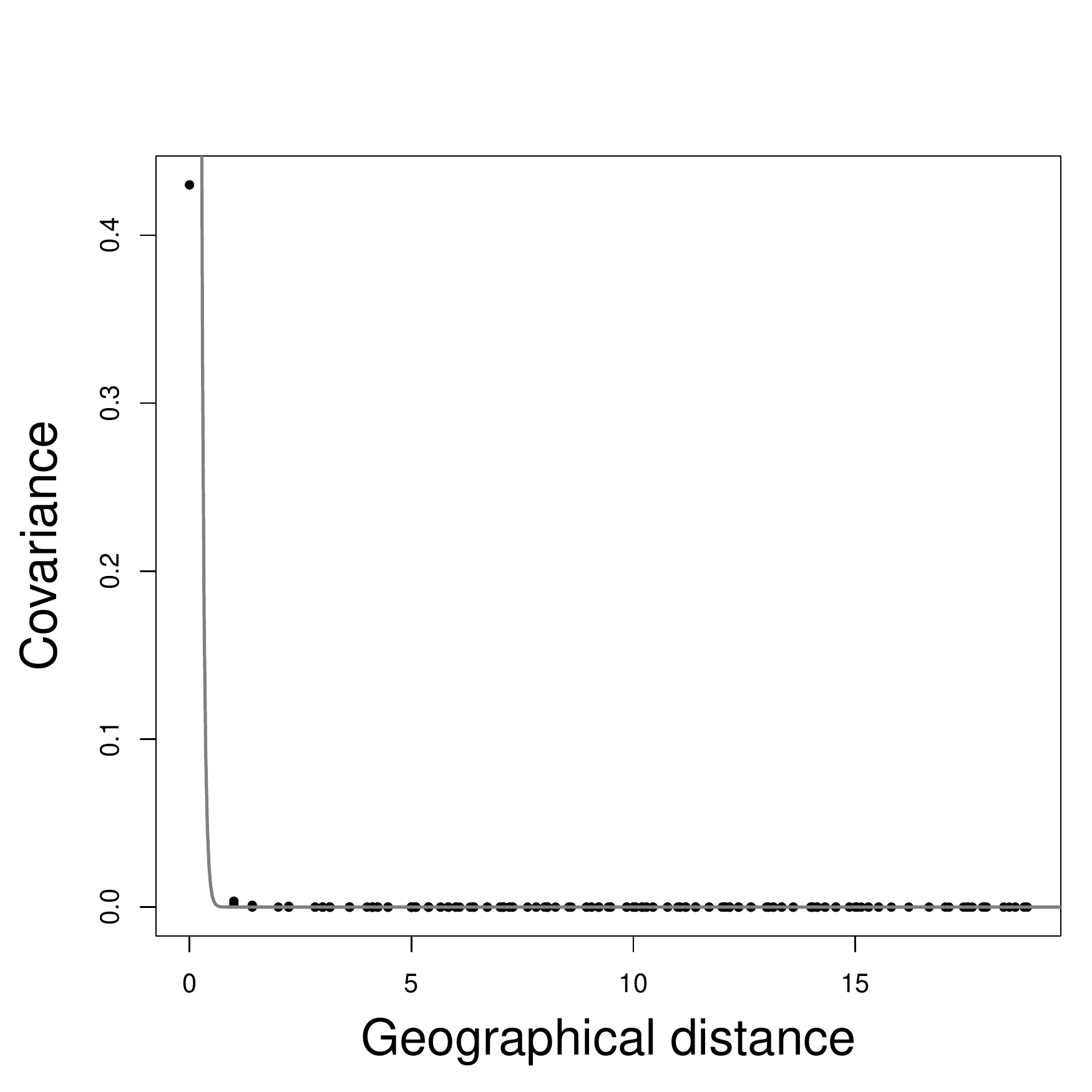}
& \vspace{-.5cm}\includegraphics[width=7.5cm]{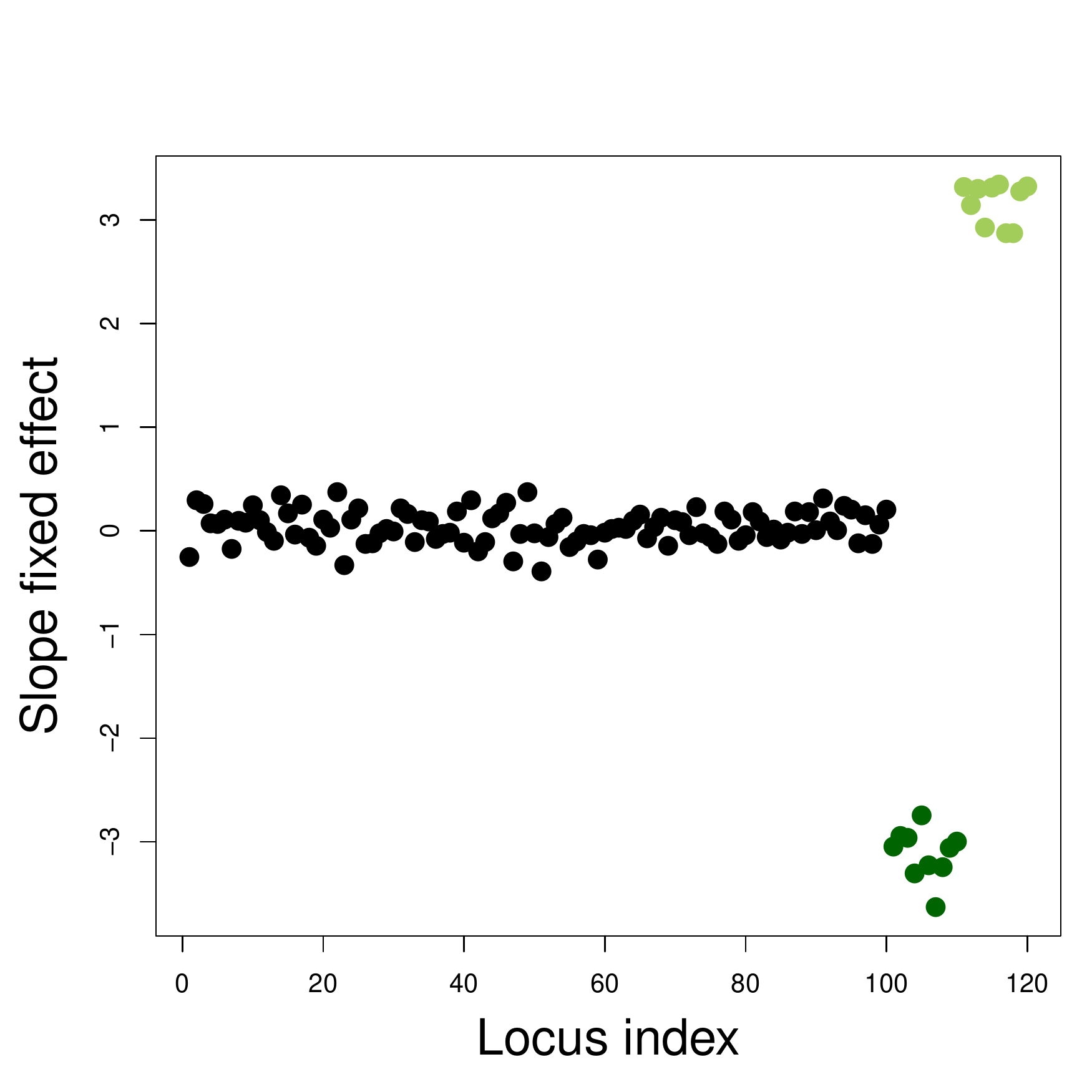}\\
 \includegraphics[width=7.5cm]{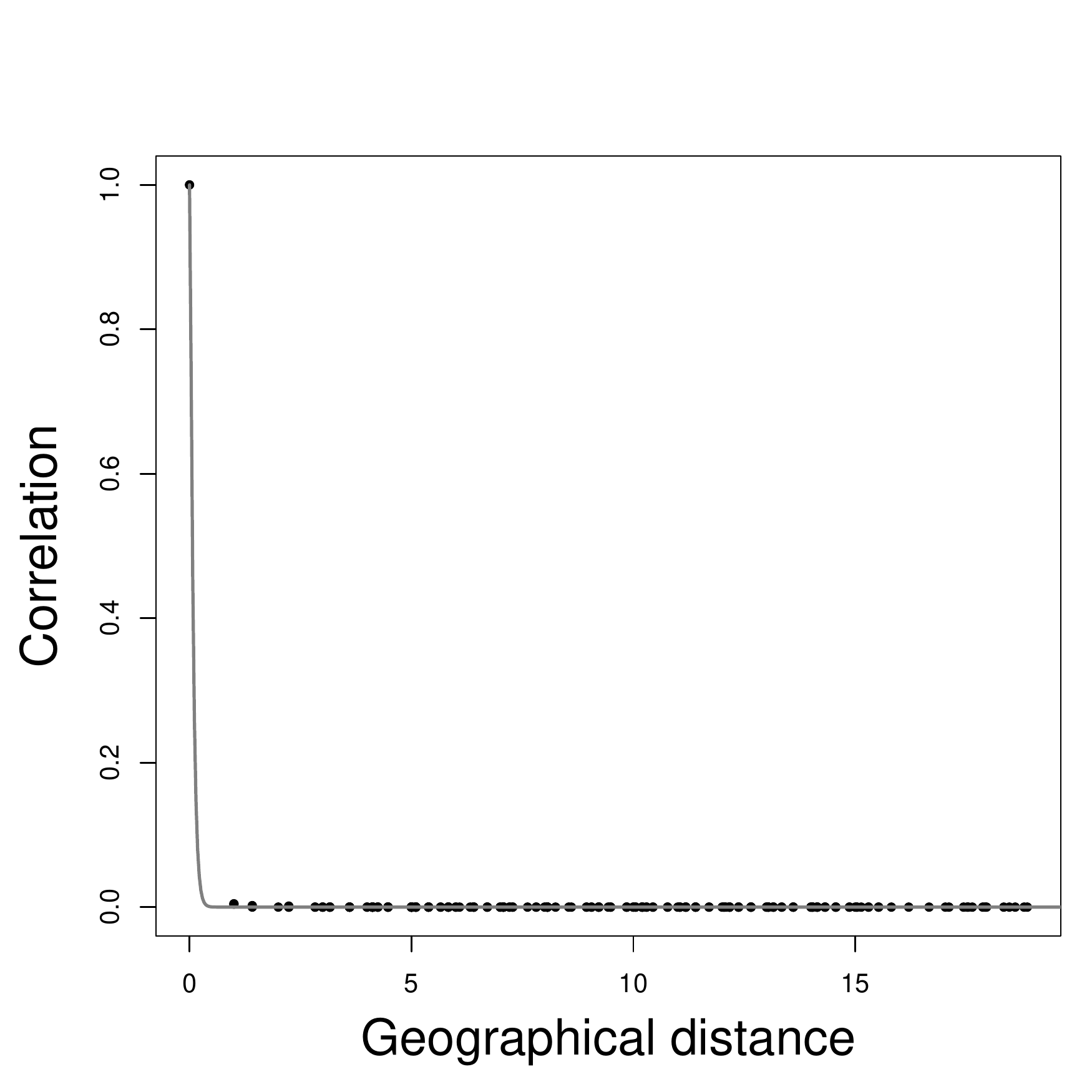} 
& \includegraphics[width=7.5cm]{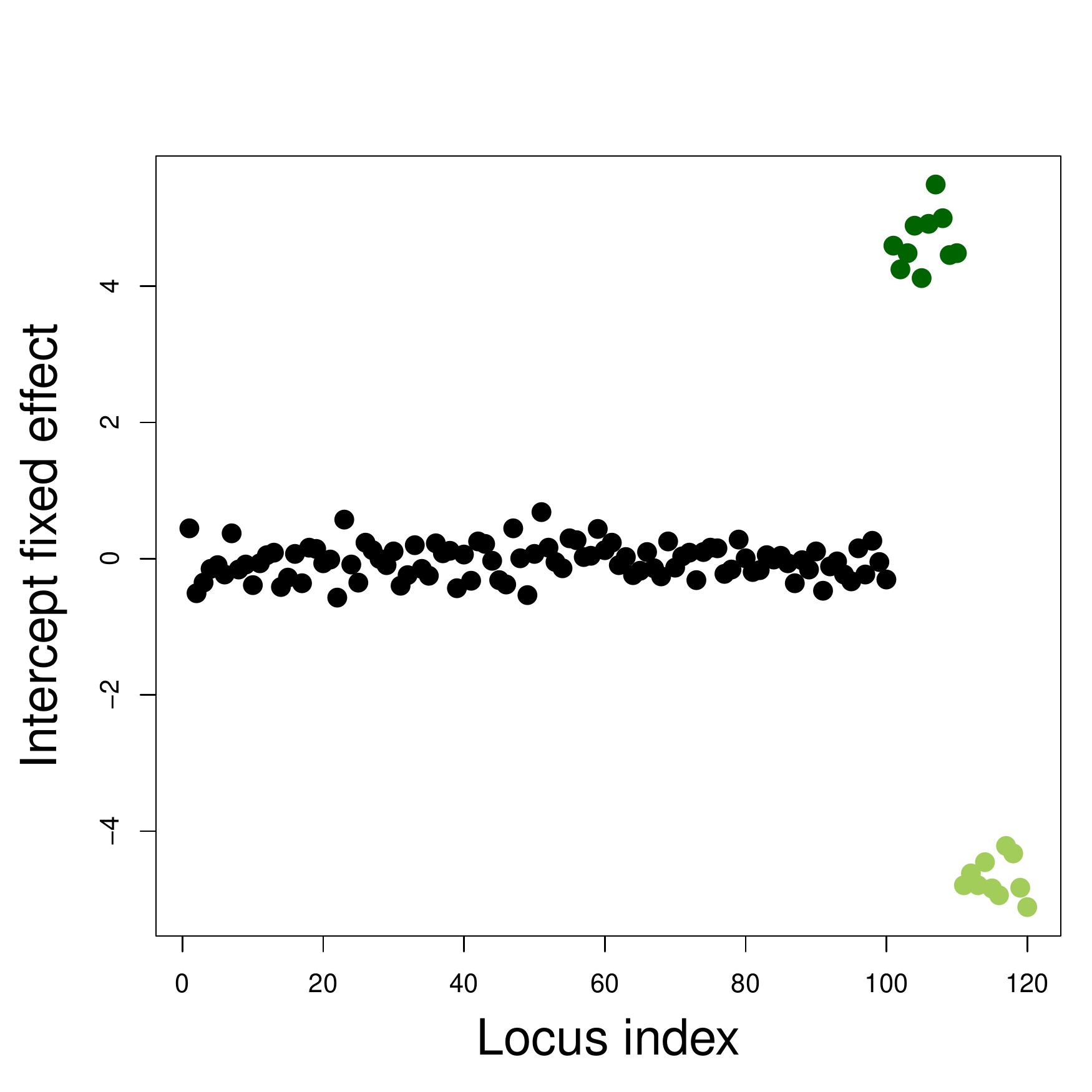}
\end{tabular}
\caption{Results for data
  simulated from a landscape genetics model (dispersal
  probability=$0.01$ per individual and per generation). 
Top left: habitat (environmental variable) coded as two colors and
sampling sites (triangles); Middle left 
  and bottom left: the   continuous grey line depicts the estimated  Mat{\'e}rn functions and the black
  dots the numerical result for the GMRF approximation underlying the
  INLA-SPDE method. 
Right from top to bottom: Bayes factors and parameters $a_l$ and $b_l$
for the 120 loci. 
Dark and light green correspond to positively and negatively selected loci respectively. 
The loci genuinely under selection are   indexed 101-120.}
\label{fig:d001}
\end{figure}

The landscape is a 30 $\times$ 10 grid of 300 cells, which can represent
habitats 1 or 2. Each cell has a carrying capacity of 100 individuals,
populations grow logistically with a rate of 0.5. The landscape is
designed so that habitats are distributed across a linear East-West
gradient of habitat frequencies (see habitat and sampling locations
figures \ref{fig:d01} and \ref{fig:d001}   top-left panel), the frequency of
habitat 1 being 1 at the eastern edge, and 0 at the western edge. The
selection coefficient is set to $s=0.1$ (each maladapted locus decreases
the fitness by 10\%). 
The probability of dispersal is set to $d=0.1$ and $d=0.01$ per
individual and per generation, 
a dispersal event consisting in moving an individual by one cell
(vertically or horizontally). 
Simulations start with a single cell at the carrying capacity (100
individuals) close to the western edge of the grid, 
and mimics the invasion of the landscape for 30 generations (enough to
reach all cells in the landscape). 
At generation 30, 25 individuals (less if the patch is not populated
enough) are sampled in each of 200 cells randomly located  in the grid, and
genotypes at both neutral and selected loci. 
Here there is a loose connection between the parametrization of our
inference model and that of the simulation model, in particular 
there is no explicit covariance function that describes the spatial
genetic structure. 
In the inference with INLA,  we use everywhere the default prior
distributions. 
What we check here is the ability of the method to detect loci that 
are genuinely under selection and its false
positive rate. 
The results are summarized in figures \ref{fig:d01} and \ref{fig:d001} 
and show good performances with respect to these two tasks. 

\clearpage
\subsection{Analysis of a pine weevil dataset in Europe}

\begin{figure}[h]
\vspace{-.0cm}\begin{tabular}{cc}
\vspace{-.2cm}\includegraphics[width=7.9cm]{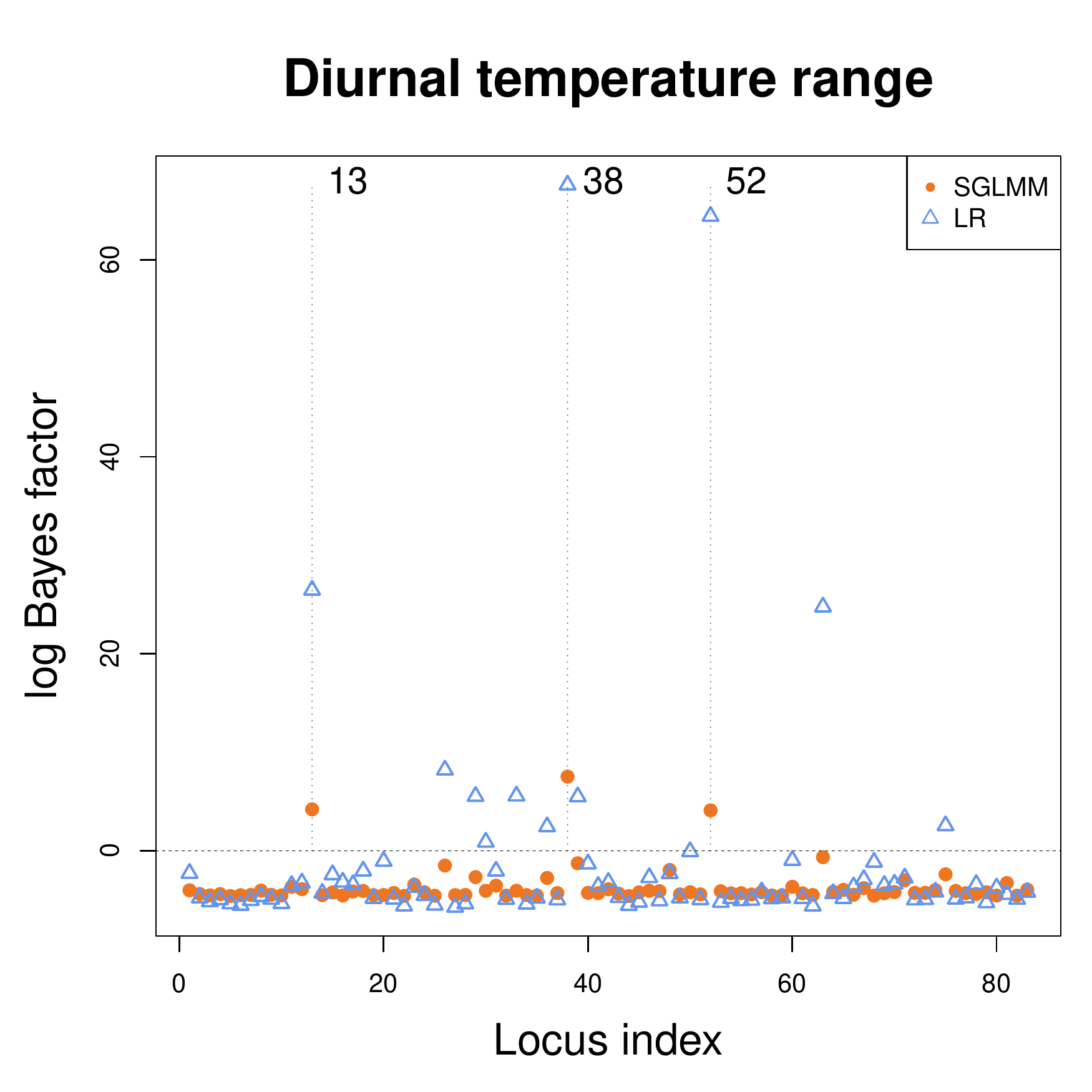}
& \vspace{-.2cm}\includegraphics[width=7.9cm]{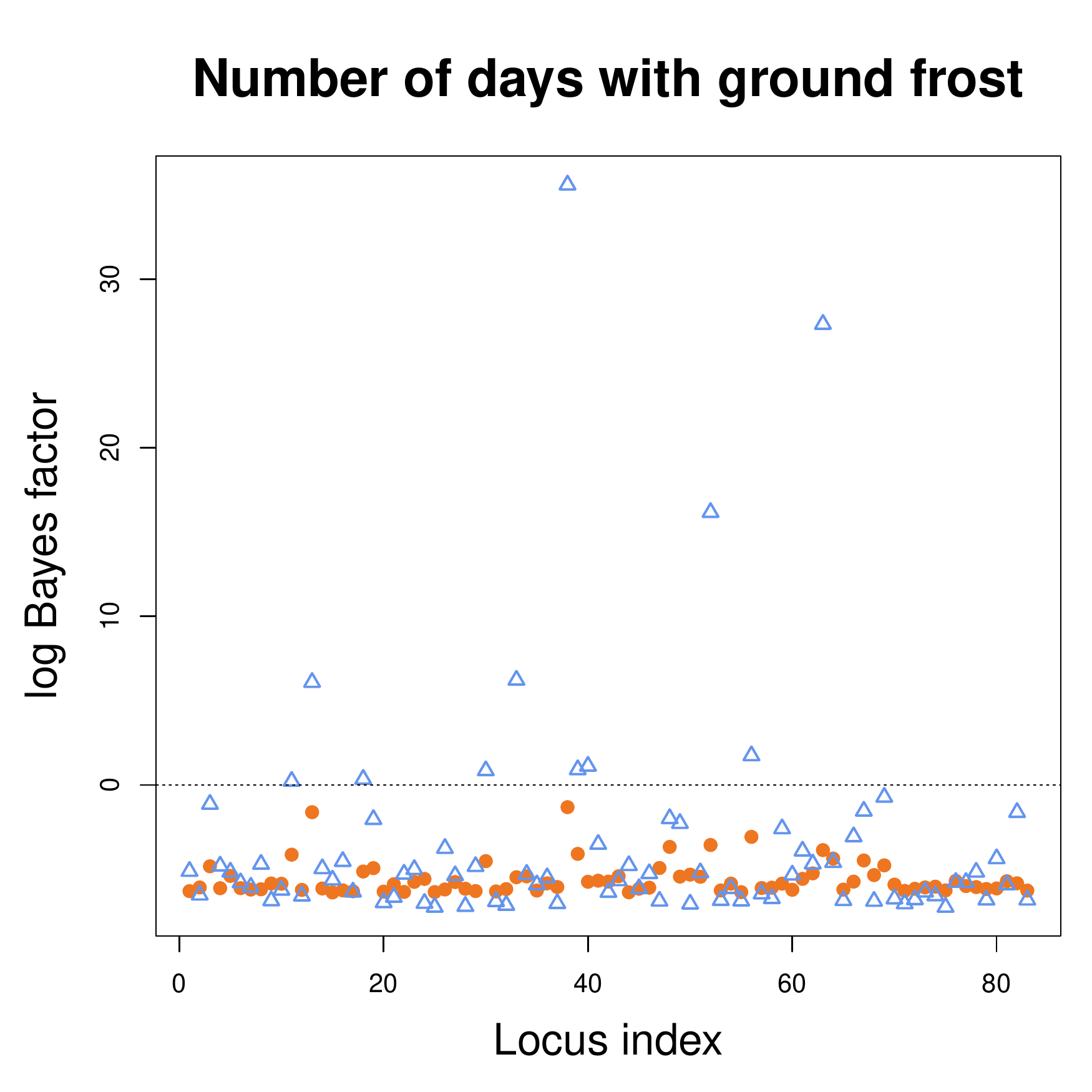}\\
\vspace{-.2cm}\includegraphics[width=7.9cm]{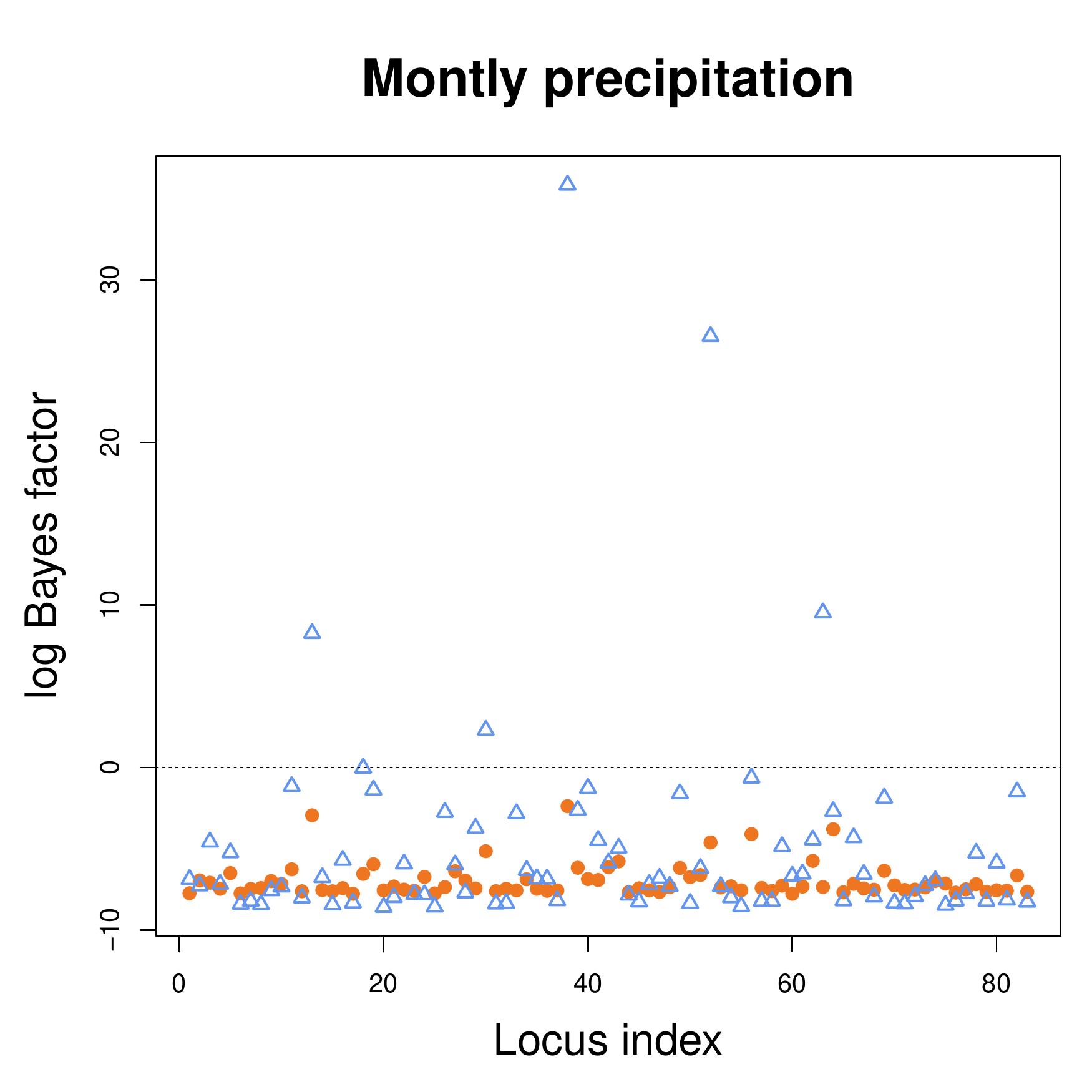}
& \vspace{-.2cm}\includegraphics[width=7.9cm]{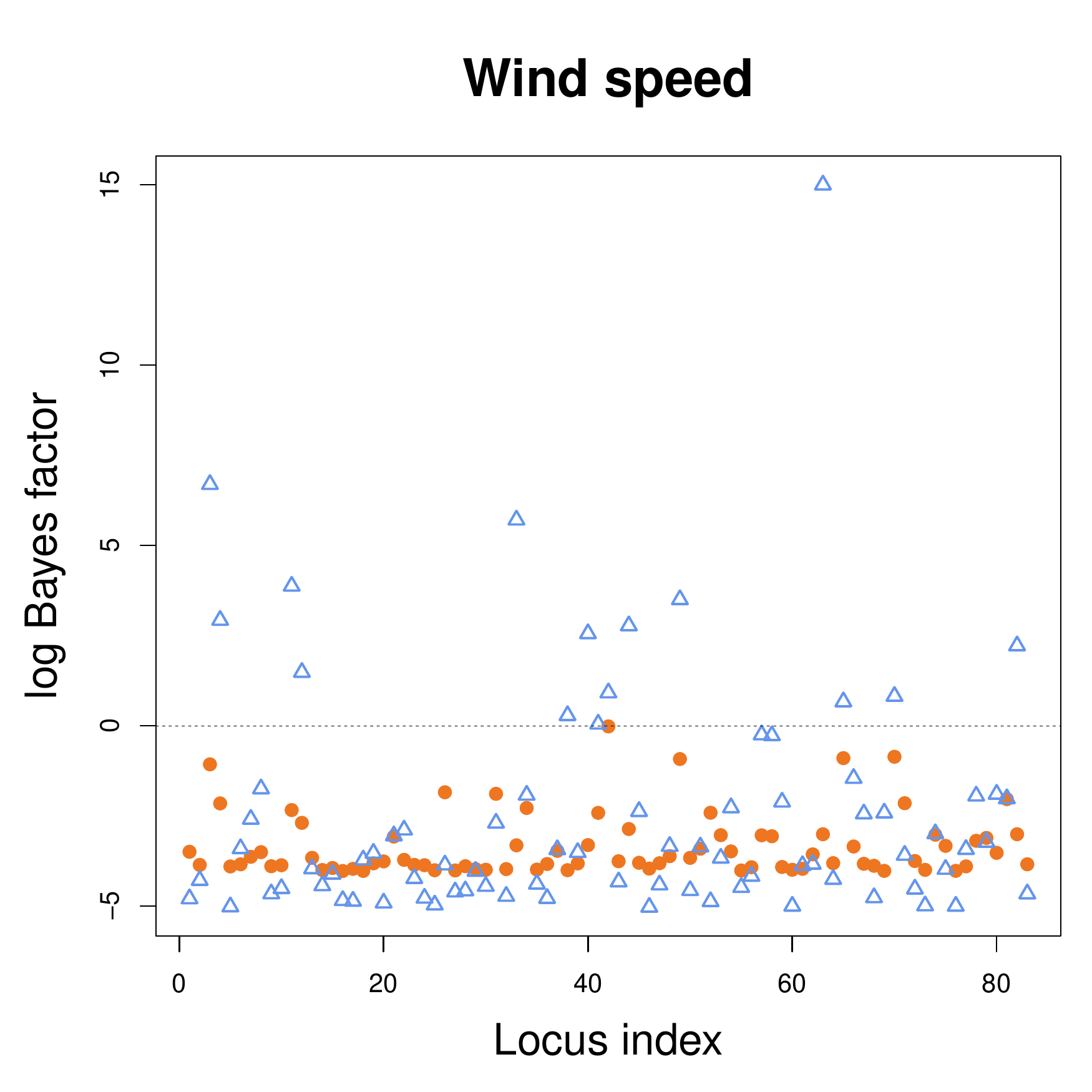}
\end{tabular}
\caption{Results of inference on the pine weevil data analyzed with a
  logistic regression (LR) and our Spatial Generalized Linear Mixed Model (SGLMM). Loci with 
  a Bayes factor in favor of a SGLMM including an effect of the
  environment variable are flagged with a vertical dot line.  }\label{fig:weevil_BF}
\end{figure}

We re-analyse this dataset with the SGLMM described above and also
with a plain logistic regression. The latter analysis differs from
that of \citet{Joost07} in the model selection strategy.
\citet{Joost07} used a somehow {\em ad hoc} combination of two tests
based 
respectively on the likelihood ratio and the Wald statistic.
 We use here Bayes factors both for the logistic regression and the
 SGLMM. 
The results for four arbitrarily chosen environmental variables out of
the ten variables  of the initial dataset are summarised in figure
\ref{fig:weevil_BF}. 
With a cut-off set at $BF>3$, under a SGLMM (resp. logistic
regression), 3 loci (resp. 11 loci)  are significantly associated with
the diurnal temperature range. 
The number of significant loci are 0 (5), 0(10) and 0(4) for the
number of days with ground frost, monthly precipitation and wind
speed. 
In the four environmental
variables considered here, only one of them is considered significantly
correlated to the genetic data under the SGLMM. 
This is consistent with the fact that the data have been collected at
highly scattered locations which makes the SGLMM better suited to
``correct'' the sample size for spatial auto-correlation. 
We note however that there is a strong agreement between the loci 
detected as most significantly associated with the environment in our
analysis under the SGLMMM and the previous analysis of \citet{Joost07}. 

\section{Conclusion}
The approach we propose extends existing methods in several ways: we introduce a method that (i) is spatially explicit, 
(ii) handles spatial coordinates either on $\mathbb{R}^2$ (plan) or
   $\mathbb{S}^2$ (sphere),  
(iii) works for both co-dominant and dominant markers,
(iv) is equally well suited for individual data or allele counts
aggregated at the population level, 
(v) does not require any calibration step on a subset of neutral loci,  
(vi)  can handle {quantitative} as well as {categorical} environmental
variables,  
(vii) returns Bayesian {measures of model fit},  
and (viii) does not rely on MCMC computation.    

One limit common to the approach proposed here and that of \citet{Coop10} 
is that loci are  assumed to be conditionally
independent (no residual linkage disequilibrium not accounted 
for by $x$ and $y$). This assumption will be clearly violated  for dense SNPs datasets. 
This aspect requires more work for a rigorous and efficient control of
false discovery. 
However we note that Bonferroni-type correction 
offers a solution to protect oneself against false positives. Moreover, potential 
linkage disequilibrium not accounted for  has no effect affect on the ranking of the loci in terms 
of evidence of selection. The approach proposed can be therefore readily used to identify conspicuous loci 
that are likely to be the target of selection. 

The model we described embeds the main features of the models of
\citet{Coop10} and the magnitude of the improvement in terms of
inference accuracy brought by the use
of an explicit spatial model depends on how much this  model
complies  with the data at hand. We expect our model to be best suited 
for datasets at a scale that is large enough to observe genetic
variation and spatial auto-correlation but small enough for  the
stationary model to make sense. The latter condition suggests that
datasets collected at the continental scale may be the best targets
for our approach.

\bigskip
\bigskip
 
\clearpage
\bibliographystyle{elsarticle-num-names.bst}

\begin{thebibliography}{14}
\providecommand{\natexlab}[1]{#1}
\providecommand{\url}[1]{\texttt{#1}}
\providecommand{\urlprefix}{URL }
\expandafter\ifx\csname urlstyle\endcsname\relax
  \providecommand{\doi}[1]{doi:\discretionary{}{}{}#1}\else
  \providecommand{\doi}[1]{doi:\discretionary{}{}{}\begingroup
  \urlstyle{rm}\url{#1}\endgroup}\fi
\providecommand{\bibinfo}[2]{#2}

\bibitem[{Rue et~al.(2009)Rue, Martino, and Chopin}]{Rue09}
\bibinfo{author}{H.~Rue}, \bibinfo{author}{S.~Martino},
  \bibinfo{author}{N.~Chopin}, \bibinfo{title}{Approximate {B}ayesian inference
  for latent {G}aussian models by using integrated nested {L}aplace
  approximations}, \bibinfo{journal}{Journal of the Royal Statistical Society,
  series B} \bibinfo{volume}{71}~(\bibinfo{number}{2}) (\bibinfo{year}{2009})
  \bibinfo{pages}{1--35}.

\bibitem[{Lindgren et~al.(2011)Lindgren, Rue, and Lindstr{\"o}m}]{Lindgren11}
\bibinfo{author}{F.~Lindgren}, \bibinfo{author}{H.~Rue},
  \bibinfo{author}{E.~Lindstr{\"o}m}, \bibinfo{title}{An explicit link between
  {G}aussian fields and {G}aussian {M}arkov random fields: the stochastic
  partial differential equation approach}, \bibinfo{journal}{Journal of the
  Royal Statistical Society, series B}
  \bibinfo{volume}{73}~(\bibinfo{number}{4}) (\bibinfo{year}{2011})
  \bibinfo{pages}{423--498}.

\bibitem[{Nielsen(2005)}]{Nielsen05}
\bibinfo{author}{R.~Nielsen}, \bibinfo{title}{Molecular signatures of natural
  selection}, \bibinfo{journal}{Annual Review of Genetics} \bibinfo{volume}{39}
  (\bibinfo{year}{2005}) \bibinfo{pages}{197--218}.

\bibitem[{Hansen et~al.(2012)Hansen, Olivieri, Waller, Nielsen, and {and the
  GEM Working Group}}]{Hansen12}
\bibinfo{author}{M.~M. Hansen}, \bibinfo{author}{I.~Olivieri},
  \bibinfo{author}{D.~M. Waller}, \bibinfo{author}{E.~E. Nielsen},
  \bibinfo{author}{{and the GEM Working Group}}, \bibinfo{title}{Monitoring
  adaptive genetic responses to environmental change},
  \bibinfo{journal}{Molecular Ecology}
  \bibinfo{volume}{21}~(\bibinfo{number}{6}) (\bibinfo{year}{2012})
  \bibinfo{pages}{1311--1329}.

\bibitem[{Yamasaki et~al.(2005)Yamasaki, Tenaillon, Vroh~{B}i, Schroeder,
  Sanchez-{V}illeda, Doebley, Gaut, and Mc{M}ullen}]{Yamasaki13}
\bibinfo{author}{M.~Yamasaki}, \bibinfo{author}{M.~I. Tenaillon},
  \bibinfo{author}{I.~Vroh~{B}i}, \bibinfo{author}{S.~Schroeder},
  \bibinfo{author}{H.~Sanchez-{V}illeda}, \bibinfo{author}{J.~Doebley},
  \bibinfo{author}{B.~Gaut}, \bibinfo{author}{M.~Mc{M}ullen}, \bibinfo{title}{A
  large scale screen for artificial selection in maize identifies candidate
  agronomic loci for domestication and crop improvement},
  \bibinfo{journal}{Plant Cell} \bibinfo{volume}{17}~(\bibinfo{number}{11})
  (\bibinfo{year}{2005}) \bibinfo{pages}{2859--2872}.

\bibitem[{Flori et~al.(2009)Flori, Fritz, Jaffrezic, Boussaha, Gut, Heath,
  Foulley, and Gautier}]{Flori09}
\bibinfo{author}{L.~Flori}, \bibinfo{author}{S.~Fritz},
  \bibinfo{author}{F.~Jaffrezic}, \bibinfo{author}{M.~Boussaha},
  \bibinfo{author}{I.~Gut}, \bibinfo{author}{S.~Heath},
  \bibinfo{author}{J.~Foulley}, \bibinfo{author}{M.~Gautier},
  \bibinfo{title}{The genome response to artificial selection: a case study in
  dairy cattl}, \bibinfo{journal}{PLoS One}
  \bibinfo{volume}{4}~(\bibinfo{number}{8}) (\bibinfo{year}{2009})
  \bibinfo{pages}{e6595}.

\bibitem[{Davey et~al.(2011)Davey, Hohenlohe, Etter, Boone, Catchen, and
  Blaxter}]{Davey11}
\bibinfo{author}{J.~Davey}, \bibinfo{author}{P.~Hohenlohe},
  \bibinfo{author}{P.~Etter}, \bibinfo{author}{J.~Boone},
  \bibinfo{author}{J.~Catchen}, \bibinfo{author}{M.~Blaxter},
  \bibinfo{title}{Genome-wide genetic marker discovery and genotyping using
  next-generation sequencing}, \bibinfo{journal}{Nature Review Genetics}
  \bibinfo{volume}{12} (\bibinfo{year}{2011}) \bibinfo{pages}{499--510}.

\bibitem[{Joost et~al.(2007)Joost, Bonin, Bruford, Despr{\'e}s, Conord,
  Erhardt, and Taberlet}]{Joost07}
\bibinfo{author}{S.~Joost}, \bibinfo{author}{A.~Bonin},
  \bibinfo{author}{M.~Bruford}, \bibinfo{author}{L.~Despr{\'e}s},
  \bibinfo{author}{C.~Conord}, \bibinfo{author}{G.~Erhardt},
  \bibinfo{author}{P.~Taberlet}, \bibinfo{title}{A spatial analysis method
  {(SAM)} to detect candidate loci for selection: towards a landscape genomics
  approach to adaptation}, \bibinfo{journal}{Molecular Ecology}
  \bibinfo{volume}{16} (\bibinfo{year}{2007}) \bibinfo{pages}{3955--3969}.

\bibitem[{Guillot and Rousset(2013)}]{Guillot13}
\bibinfo{author}{G.~Guillot}, \bibinfo{author}{F.~Rousset},
  \bibinfo{title}{Dismantling the {M}antel tests}, \bibinfo{journal}{Methods in
  Ecology and Evolution} \bibinfo{volume}{4}~(\bibinfo{number}{4})
  (\bibinfo{year}{2013}) \bibinfo{pages}{336--344}.

\bibitem[{Coop et~al.(2010)Coop, Witonsky, {D}i {R}ienzo, and
  Pritchard}]{Coop10}
\bibinfo{author}{G.~Coop}, \bibinfo{author}{D.~Witonsky},
  \bibinfo{author}{A.~{D}i {R}ienzo}, \bibinfo{author}{J.~Pritchard},
  \bibinfo{title}{Using environmental correlations to identify loci under
  selection}, \bibinfo{journal}{Genetics} \bibinfo{volume}{185}
  (\bibinfo{year}{2010}) \bibinfo{pages}{1411--1423}.

\bibitem[{De~Mita et~al.(2013)De~Mita, Thuillet, Gay, Ahmadi, Manel, Ronfort,
  and Vigouroux}]{deMita13}
\bibinfo{author}{S.~De~Mita}, \bibinfo{author}{A.~Thuillet},
  \bibinfo{author}{L.~Gay}, \bibinfo{author}{N.~Ahmadi},
  \bibinfo{author}{S.~Manel}, \bibinfo{author}{J.~Ronfort},
  \bibinfo{author}{Y.~Vigouroux}, \bibinfo{title}{Detecting selection along
  environmental gradients: analysis of eight methods and their effectiveness
  for outbreeding and selfing populations}, \bibinfo{journal}{Molecular
  Ecology} \bibinfo{volume}{22}~(\bibinfo{number}{5}) (\bibinfo{year}{2013})
  \bibinfo{pages}{1383--1399}.

\bibitem[{Conord et~al.(2006)Conord, Lemp\'eri\`ere, Taberlet, and
  Despr\'es}]{Conord06}
\bibinfo{author}{C.~Conord}, \bibinfo{author}{G.~Lemp\'eri\`ere},
  \bibinfo{author}{P.~Taberlet}, \bibinfo{author}{L.~Despr\'es},
  \bibinfo{title}{Genetic structure of the forest pest {\em {H}ylobius abietis}
  on conifer plantations at different spatial scales in Europe},
  \bibinfo{journal}{Heredity} \bibinfo{volume}{97} (\bibinfo{year}{2006})
  \bibinfo{pages}{46--55}.

\bibitem[{Chil{\`e}s and Delfiner(1999)}]{Chiles99}
\bibinfo{author}{J.~Chil{\`e}s}, \bibinfo{author}{P.~Delfiner},
  \bibinfo{title}{Geostatistics: Modeling Spatial Uncertainty},
  \bibinfo{publisher}{Wiley}, \bibinfo{address}{Hoboken, NJ, USA},
  \bibinfo{year}{1999}.

\bibitem[{Rebaudo et~al.(2013)Rebaudo, Le~Rouzic, Dupas, Silvain, Harry, and
  Dangles}]{Rebaudo13}
\bibinfo{author}{F.~Rebaudo}, \bibinfo{author}{A.~Le~Rouzic},
  \bibinfo{author}{S.~Dupas}, \bibinfo{author}{J.~Silvain},
  \bibinfo{author}{M.~Harry}, \bibinfo{author}{O.~Dangles},
  \bibinfo{title}{SimAdapt: An individual-based genetic model for simulating
  landscape management impacts on populations}, \bibinfo{journal}{Methods in
  Ecology and Evolution} \bibinfo{volume}{4}~(\bibinfo{number}{6})
  (\bibinfo{year}{2013}) \bibinfo{pages}{595--600}.

\end{thebibliography}

\appendix
\newpage







\end{document}